\documentclass[a4paper,11pt]{article}
\usepackage{jcappub} 
\usepackage{amsmath,bm}
\usepackage{lineno}
\usepackage{xcolor}

\title{\boldmath SASHIMI-SIDM: Semi-analytical subhalo modelling for self-interacting dark matter at sub-galactic scales}

\author[a,b]{Shin'ichiro Ando,}
\author[b]{Shunichi Horigome,}
\author[c,d,e]{Ethan O. Nadler,}
\author[f]{Daneng Yang}
\author[f]{and Hai-Bo Yu}

\affiliation[a]{GRAPPA Institute, University of Amsterdam, Science Park 904, 1098 XH Amsterdam, The Netherlands}
\affiliation[b]{Kavli Institute for the Physics and Mathematics of the Universe, University of Tokyo, Chiba 277-8583, Japan}
\affiliation[c]{Department of Astronomy \& Astrophysics, University of California, San Diego, La Jolla, CA 92093, USA}
\affiliation[d]{Carnegie Observatories, 813 Santa Barbara Street, Pasadena, CA 91101, USA}
\affiliation[e]{Department of Physics \& Astronomy, University of Southern California, Los Angeles, CA 90007, USA}
\affiliation[f]{Department of Physics and Astronomy, University of California, Riverside, CA 92521, USA}

\emailAdd{s.ando@uva.nl}

\abstract{We combine the semi-analytical structure formation model, SASHIMI, which predicts subhalo populations in collisionless, cold dark matter (CDM), with a parametric model that maps CDM halos to self-interacting dark matter (SIDM) halos. The resulting model, SASHIMI-SIDM, generates SIDM subhalo populations down to sub-galactic mass scales, for an arbitrary input cross section, in minutes. We show that SASHIMI-SIDM agrees with SIDM subhalo populations from high-resolution cosmological zoom-in simulations in resolved regimes. Crucially, we predict that the fraction of core-collapsed subhalos peaks at a mass scale determined by the input SIDM cross section and decreases toward higher halo masses, consistent with the predictions of gravothermal models and cosmological simulations. {For the first time, we also show that the core-collapsed fraction decreases toward lower halo masses. While the dependence of the collapse time on mass and concentration implies such behaviour, our semi-analytical approach allows us to quantify and illustrate this trend clearly across the full mass spectrum of subhalos, including for subhalo masses below the resolution limit of any current cosmological SIDM simulation.} As a proof of principle, we apply SASHIMI-SIDM to predict the boost to the local dark matter density and annihilation rate from core-collapsed SIDM subhalos, which can be enhanced relative to CDM by an order of magnitude for viable SIDM models. Thus, SASHIMI-SIDM provides an efficient and reliable tool for scanning SIDM parameter space and testing it with astrophysical observations. The code is publicly available at \url{https://github.com/shinichiroando/sashimi-si}.}

\begin{document}
\maketitle
\flushbottom

\section{Introduction}
\label{sec:intro}

Dark matter accounts for more than 80\% of the total matter in the Universe. Yet, its particle identity remains unknown, and understanding its properties as an elementary particle is a pressing task in physics. Whether the dark matter particles have non-gravitational interactions is one of the most intriguing questions one can tackle using modern astrophysical observations. In fact, various observations at sub-galactic scales indicate that the collisionless cold dark matter (CDM) scenario may not be complete, and that new physics may be needed, see, e.g., \cite{Bullock:2017xww,Nesti:2023tid}. As a compelling solution, self-interacting dark matter (SIDM), which was originally proposed to address the tensions between CDM predictions and observations of galactic systems~\cite{Spergel:1999mh}, has attracted broad attention from multiple communities of particle physics, astrophysics, and cosmology; see \cite{Tulin:2017ara,Adhikari:2022sbh} for recent reviews.

In the SIDM scenario, dark matter particles can scatter with each other and exchange momentum. At the first stage of the gravothermal evolution of an SIDM halo, the heat is transported from the hotter outer region to the colder inner region, resulting in a flat density core towards the center, in contrast to a density cusp in CDM~\cite{Navarro:1995iw}. At the later stage, the inner region becomes hotter and the self-interactions transport heat outwards, leading to a high central density --- a phenomenon called `gravothermal collapse'~\cite{Balberg:2002ue,Koda:2011yb}. The collapse timescale depends on the self-interacting cross section per particle mass and the halo parameters, particularly, the halo concentration~\cite{Essig:2018pzq,Kaplinghat:2019svz,Nadler:2023nrd}. Therefore, SIDM predicts more diverse dark matter distributions in galaxies than CDM~\cite{Kamada:2016euw,Ren:2018jpt,Sameie:2019zfo,Correa:2020qam,Turner:2020vlf,Correa:2022dey,Yang:2023jwn,Nadler:2023nrd}. For satellite halos, environmental effects such as tidal stripping could further accelerate the onset of the collapse by lowering the temperature height of the halo~\cite{Nishikawa:2019lsc,Sameie:2019zfo,Kahlhoefer:2019oyt,Zeng:2021ldo,Yang:2021kdf,Shirasaki:2022ttb,Andrade:2023fgr}. Thus, there is a novel interplay between gravitational evolution and tidal evolution for subhalos.

N-body simulations are leading tools to make SIDM predictions; see, e.g., \cite{Meskhidze:2022hwm,Palubski:2024ibb,Fischer:2024eaz} for recent work on subtleties of SIDM implementations in numerical simulations. Nevertheless, they are typically computationally expensive and less flexible in exploring a broad range of SIDM parameter space. 
Recently, Ref.~\cite{Yang:2023jwn} proposed a parametric model for modelling SIDM halos and it is a semi-analytical method that can map CDM (sub)halos to their SIDM counterparts for a given particle physics realisation of SIDM. 
{The model builds on a body of recent work demonstrating that the gravothermal evolution of the bulk regions of SIDM halos is universal~\cite{Outmezguine:2022bhq,Yang:2023jwn,Zhong:2023yzk,Yang:2024tba}. It uses an effective constant cross section to simplify the underlying differential SIDM cross section~\cite{Yang:2022hkm,Yang:2022zkd}.   Moreover, an integral approach is introduced to incrementally incorporate SIDM effects along the evolutionary histories of CDM. 
Ref.~\cite{Yang:2024uqb} has extensively tested the model against halos and subhalos in cosmological SIDM zoom-in simulations from Refs.~\cite{Yang:2022mxl,Nadler:2023nrd}, finding that it accurately predicts subhalo density profiles, on a subhalo-by-subhalo basis, at the $10\%$ to $50\%$ level. }
In this work, we further implement the parametric model into a semi-analytical tool for subhalo populations, \textbf{S}emi-\textbf{A}nalytical \textbf{S}ub\textbf{H}alo \textbf{I}nference \textbf{M}odell\textbf{I}ng (SASHIMI).\footnote{See \url{https://github.com/shinichiroando/sashimi-c} for the CDM version of SASHIMI.} SASHIMI is based on the theory of structure formation and it can compute CDM subhalo populations efficiently, avoiding issues related to numerical resolution and shot noise inevitable for simulation-based approaches. Being in good agreement with outcomes of cosmological CDM simulations such as subhalo mass functions~\cite{Hiroshima:2018kfv} and distributions of subhalo density profile parameters~\cite{Ando:2020yyk}, SASHIMI provides a reliable way to extend simulation results into sub-resolution regimes.

Here, we develop a new semi-analytical tool, SASHIMI-SIDM, for semi-analytically modelling SIDM subhalo populations. We will demonstrate that SASHIMI-SIDM predictions agree well with those from high-resolution cosmological SIDM simulations. For low-mass subhalos, one of our main findings is that the fraction of subhalos that are core-collapsed peaks at a mass scale determined by the SIDM cross section and decreases towards both higher and lower subhalo masses. As an application, we will evaluate the enhancement factors due to the collapse of SIDM subhalos for both direct and indirect detection experiments. For indirect detection, we find a boost for the rate of dark matter annihilation by up to an order of magnitude, with details depending on the SIDM model. Our method generates subhalo populations to be computed within minutes for a given host halo, will enable efficient scans over SIDM parameter space, and can be used to test regions of SIDM parameter space in a continuous manner. SASHIMI-SIDM is publicly available at \url{https://github.com/shinichiroando/sashimi-si}.

The rest of the paper is organised as follows.
In section~\ref{sec:Semi-analytical modelling of SIDM subhalos}, we describe SASHIMI-SIDM by going over each model ingredient and discuss the model validity at sub-resolution scales. In section~\ref{sec:Results}, we show various SASHIMI-SIDM outputs, including density profiles of cored and collapsed subhalos, subhalo mass function, and distribution of the density profile parameters. Furthermore, we validate our results using the corresponding simulation outputs, and then predict the subhalo populations in the sub-resolution regime. In section~\ref{sec:Discussion}, we discuss applications of the subhalo populations with SASHIMI-SIDM for both direct and indirect detection experiments, and the dependence on SIDM models. We conclude in section~\ref{sec:Conclusions}. Throughout the paper, we adopt the following cosmological parameters: $H_0 = 100h$~km~s$^{-1}$~Mpc$^{-1}$ with $h = 0.674$, $\Omega_m = 0.315$, and $\Omega_\Lambda = 0.685$. We use `$\ln$' to represent the natural logarithm, while `$\log$' is used for 10-based logarithm.

\section{Semi-analytical modelling of SIDM subhalos}
\label{sec:Semi-analytical modelling of SIDM subhalos}

\subsection{SASHIMI for CDM}
\label{sec:SASHIMI for CDM}

SASHIMI is a publicly available numerical code for calculating various subhalo properties using semi-analytical models of structure formation~\cite{Hiroshima:2018kfv}.
SASHMI computes the properties of CDM subhalos semi-analytically without being hindered by numerical resolutions or statistical noise.
Given a host specified with its mass, redshift, and concentration parameter, it combines the accretion of CDM subhalos onto the host based on the extended Press-Schechter (ePS) formalism~\cite{Bond:1990iw} with the modelling of subhalo evolution after accretion onto a host halo.

By adopting the ePS formalism, which has been well calibrated against the numerical simulations (e.g., \cite{Yang:2011rf}), we obtain the average progenitor mass function per unit redshift $z_a$ per unit subhalo mass $m_a$, ${d^2N_{\rm sh}}/{(dm_adz_a)}$.
This quantity is represented as a function of (linearly extrapolated) critical over-density $\delta_c(z)$ of halo collapse at a given redshift $z$ and mass variance $\sigma^2(m)$ for halo and subhalo masses $m$. We refer the reader to refs.~\cite{Yang:2011rf, Hiroshima:2018kfv} for detailed expressions of subhalo accretion rate $d^2N_{\rm sh}/(dm_adz_a)$. {In this work, we sample the average progenitor mass function in order to obtain statistical predictions for ensembles of subhalo populations. It will be interesting to extend our framework to build individual merger trees and thus model halo-to-halo variance, following Ref.~\cite{Hiroshima:2022khy}.}

Once the subhalo with the mass $m_{a}$ accretes onto its host at $z_a$, it loses mass via tidal stripping.
The following differential equation describes the mass evolution:
\begin{equation}
    \dot m(z) = -A \frac{m(z)}{\tau_{\rm dyn}(z)}\left[\frac{m(z)}{M(z)}\right]^\zeta,
    \label{eq:subhalo mass loss}
\end{equation}
where $A$ and $\zeta$ are two parameters fitted against Monte Carlo simulations that are found to vary weakly as functions of $M$ and $z$~\cite{Hiroshima:2018kfv} {(see Refs.~\cite{vandenBosch:2004zs, Jiang:2014nsa} for earlier work)}.
For the redshift evolution of the host mass $M(z)$, we adopt the result of ref.~\cite{Correa:2014xma}.
We solve eq.~(\ref{eq:subhalo mass loss}) with the initial condition of $m(z_a) = m_a$ to obtain the subhalo mass $m$ at any redshift $z$ between $z_a$ and 0.

While the subhalo loses mass, its density profile is described by the Navarro-Frenk-White (NFW) function~\cite{Navarro:1995iw}:
\begin{equation}
    \rho_{\rm CDM}(r) = \frac{\rho_s}{(r/r_s)(r/r_s+1)^2},
    \label{eq:NFW}
\end{equation}
where $\rho_s$ is the density profile normalisation, $r_s$ is the scale radius, and the profile is cut off abruptly beyond the tidal truncation radius $r_t$, which is obtained from mass conservation:
\begin{equation}
    m = 4\pi \int_0^{r_t} dr r^2 \rho_{\rm CDM}(r) = 4\pi \rho_s r_s^3f\left(\frac{r_t}{r_s}\right).
\end{equation}

We model the changes of $r_s$ and $\rho_s$ in terms of the related parameters $r_{\rm max}$ and $V_{\rm max}$,
\begin{eqnarray}
    r_s &=& \frac{r_{\rm max}}{2.163},\label{eq:r_s-r_max}
    \\
    \rho_s &=& \frac{4.625}{4\pi G}\left(\frac{V_{\rm max}}{r_s}\right)^2,\label{eq:rho_s-V_max}
\end{eqnarray}
where $V_{\rm max}$ is the maximum circular velocity and $r_{\rm max}$ is the radius at which $V_{\rm max}$ is reached. 
The initial values of $r_s$ and $\rho_s$ at accretion are obtained via
\begin{eqnarray}
    m_{a} &=& \frac{4\pi}{3}\Delta_c(z_a)\rho_c(z_a)r_{{\rm vir},a}^3, \\
    r_{s, a} &=& \frac{r_{{\rm vir}, a}}{c_{a}},\\
    \rho_{s,a} &=& \frac{m_a}{4\pi r_{s, a}^3 f(c_{{\rm vir},a})},
\end{eqnarray}
where $\Delta_c = 18\pi^2+82 d-39 d^2$~\cite{Bryan:1997dn}, $d = \Omega_m(1+z_a)^3/[\Omega_m(1+z_a)^3+\Omega_\Lambda]-1$, $\rho_c(z_a)$ is the critical density of the Universe at $z_a$, and $f(c) = \ln(1+c)-c/(1+c)$.
The virial concentration parameter $c_{a}$ is generated following the log-normal distribution function with the mean $\bar c_{\rm vir}(m_a,z_a)$ from refs.~\cite{Correa:2014xma,Correa:2015dva}, with appropriate conversions based on our halo mass definitions and a scatter of $\sigma_{\log c} = 0.13$~\cite{Ishiyama:2011af}.

Following ref.~\cite{Penarrubia:2010jk}, we parameterise the evolution of $r_{\rm max}$ and $V_{\rm max}$ after accretion as
\begin{eqnarray}
    \frac{V_{\rm max}}{V_{{\rm max},a}} &=& \frac{2^{0.4}(m/m_a)^{0.3}}{(1+m/m_a)^{0.4}}, \\
    \frac{r_{\rm max}}{r_{{\rm max},a}} &=& \frac{2^{-0.3}(m/m_a)^{0.4}}{(1+m/m_a)^{-0.3}}.
\end{eqnarray}
Using eqs.~(\ref{eq:r_s-r_max}) and (\ref{eq:rho_s-V_max}) again, we obtain $r_s$ and $\rho_s$ at an arbitrary redshift $0<z<z_0$.

\subsection{Parametric model of SIDM}
\label{sec:Parametric model of SIDM}

{The parametric model of SIDM introduced in \cite{Yang:2023jwn} connects parameters of an SIDM subhalo to the corresponding one in CDM. Ref.~\cite{Yang:2024uqb} demonstrates that it accurately predicts SIDM subhalo density profiles, on a subhalo-by-subhalo basis, when compared to the cosmological zoom-in simulation of our Model~I, from Ref.~\cite{Yang:2022mxl}, and our Model~III, from Ref.~\cite{Nadler:2023nrd}. 
Once we know the evolution history of CDM subhalos and their density profile parameters, the parametric model allows us to compute the SIDM counterpart following the prescription detailed in this subsection. The parametric model accurately captures SIDM subhalos' tidal evolution because SIDM primarily alters the inner halo structure, while tidal stripping starts from the outer regions and progressively moves inward. Consequently, most surviving subhalos have mass evolution histories similar to their CDM counterparts, and the primary effects of SIDM on these systems can be captured through the evolution of $V_{\rm max}$ and $R_{\rm max}$ (see Ref.~\cite{Yang:2024uqb} for further discussion). Given the difficulties of tracking disrupting, cored subhalos in cosmological simulations, we conservatively assume that every CDM subhalo maps to an SIDM subhalo (i.e., we neglect full tidal disruption) to provide an upper limit on the expected subhalo populations. For Model~I, this assumption is accurate based on the similarity between CDM and SIDM subhalo mass functions reported by Ref.~\cite{Nadler:2023nrd}. Note that our key results concern the fraction of core-collapsed halos, which are resilient to tidal disruption, and are thus unaffected by this assumption.}

To apply the parametric SIDM model to our SASHIMI predictions, we adopt the same ePS formalism $d^2N_{\rm sh}/(dm_adz_a)$ as in the case of CDM. 
{(Likewise, we assume that the concentration-mass relation and the formation redshift-halo mass relation are unchanged due to the SIDM effect.)}
This is based on two assumptions: ($i$) the linear power spectrum is unchanged between CDM and SIDM (i.e., we only consider ``late-time'' effects of SIDM), and ($ii$) self-interactions generally have a minor impact on halo abundances and orbital properties before infall~\cite{Nadler:2020ulu,Yang:2022mxl}. The second assumption should be revisited carefully for models that reach very large cross sections (e.g.,~\cite{Nadler:2023nrd}). {Note that our calculation of orbit-averaged subhalo properties in CDM commutes with our application of the parametric SIDM model. Future work that models subhalos' spatial and orbital distributions will need to revisit this assumption.}

For the density profile of the SIDM subhalos, we adopt the following parameterisation:
\begin{equation}
    \rho_{\rm SIDM}(r) = \frac{\rho_s}{[{(r^\beta+r_c^\beta)^{1/\beta}}/{r_s}](r/r_s+1)^2},
    \label{eq:density profile SIDM}
\end{equation}
up to the tidal truncation radius $r_t$, beyond which the density decreases quickly (for a more accurate treatment of the cutoff, see~\cite{Yang:2023jwn}). 
In eq.~(\ref{eq:density profile SIDM}), $r_c$ is the core radius within which the density profile is flat.
We adopt $\beta = 4$ throughout this work.
Instead of directly using $(\rho_s, r_s,r_c)$, we work in the basis of $(V_{\rm max}, r_{\rm max})$ and then convert to these parameters.
Following the integral model approach of ref.~\cite{Yang:2023jwn}, these parameter sets of SIDM can be obtained from the density profile parameters of CDM through the relation:
\begin{eqnarray}
    V_{\rm max}^{\rm SIDM}(t) &=& V_{\rm max}^{\rm CDM}(t)+\int_{t_f}^t \frac{dt'}{t_c(t')}\frac{dV_{\rm max,Model}(\tilde t')}{d\tilde t'}, 
    \label{eq:V_max}\\
    r_{\rm max}^{\rm SIDM}(t) &=& r_{\rm max}^{\rm CDM}(t)+\int_{t_f}^t \frac{dt'}{t_c(t')}\frac{dr_{\rm max,Model}(\tilde t')}{d\tilde t'},
    \label{eq:rmax}\\
    \frac{dV_{\rm max,Model}(\tilde t)}{d\tilde t} &=& V_{\rm max}^{\rm CDM}(t_f) \left[0.1777-4.399(3\tilde t^2)+16.66(4\tilde t^3)-18.87(5\tilde t^4) \right.\nonumber\\&&{}\left.
    +9.077(7\tilde t^6)-2.436(9\tilde t^8) \right]\Theta(1.1-\tilde t), \\
    \frac{dr_{\rm max,Model}(\tilde t)}{d\tilde t} &=& r_{\rm max}^{\rm CDM}(t_f) \left[0.007623-0.7200(2\tilde t)+0.3376(3\tilde t^2)-0.1375(4\tilde t^3)\right]\Theta(1.1-\tilde t),\nonumber\\
    \label{eq:r_c}
\end{eqnarray}
where $t_f$ is the time when the given subhalo is formed, $\tilde t = (t-t_f)/t_c(t)$, $t_c(t)$ is the core-collapse timescale (defined below), and $\Theta$ is the Heaviside step function.
The second term of the right-hand side of eqs.~(\ref{eq:V_max}) and (\ref{eq:rmax}) takes the SIDM effects into account such as core formation and collapse, while the first term is obtained as a result of the tidal effects.
By using the step function $\Theta$, we terminate the SIDM effect (i.e., the second term) once the collapse phase has been reached, $\tilde t > 1.1$, up to which the parametric model has been calibrated against N-body simulations~\cite{Yang:2023jwn}. {Note that this assumption is only relevant for extreme, deeply core-collapsed subhalos, and thus does not significantly affect our key findings, particularly in the very low-mass subhalo regime.}

The conversion from $(V_{\rm max}, r_{\rm max})$ to $(\rho_s, r_s, r_c)$ for SIDM subhalos can be implemented as follows.
Firstly, starting with $V_{\rm max}^{\rm SIDM}(t)$ and $r_{\rm max}^{\rm SIDM}(t)$ obtained above, we compute $V_{\rm max,0}^{\rm CDM}$ and $r_{\rm max,0}^{\rm CDM}$ using the following relations:
\begin{eqnarray}
\frac{V_{\rm max}^{\rm SIDM}(t)}{V_{\text{max,0}}^{\rm CDM}} &=& 1 + 0.1777 \tilde{t} - 4.399 \tilde{t}^3 + 16.66 \tilde{t}^4 - 18.87 \tilde{t}^5 + 9.077 \tilde{t}^7 - 2.436 \tilde{t}^9, 
\label{eq:Vmax0}\\
\frac{r_{\rm max}^{\rm SIDM}(t)}{r_{\text{max,0}}^{\rm CDM}} &=& 1 + 0.007623 \tilde{t} - 0.7200 \tilde{t}^2 + 0.3376 \tilde{t}^3 - 0.1375 \tilde{t}^4.
\label{eq:rmax0}
\end{eqnarray}
For the subhalos which have reached the collapsing phase, we adopt $\tilde t = 1.1$.
$V_{\rm max,0}^{\rm CDM}$ and $r_{\rm max,0}^{\rm CDM}$ are the initial values of these parameters of a {\it fictitious} CDM halo.
Secondly, we convert ($V_{\rm max,0}^{\rm CDM}$, $r_{\rm max,0}^{\rm CDM}$) to $(\rho_{s,0}^{\rm CDM}, r_{s,0}^{\rm CDM})$ using eqs.~(\ref{eq:r_s-r_max}) and (\ref{eq:rho_s-V_max}) for the same fictitious CDM halo.
Finally, using 
\begin{eqnarray}
    \frac{\rho_s^{\rm SIDM}(t)}{\rho_{s,0}^{\rm CDM}} &=& 2.033 + 0.7381 \tilde{t} + 7.264 \tilde{t}^5 - 12.73 \tilde{t}^7 + 9.915 \tilde{t}^9
    + (1 - 2.033) \frac{\ln (\tilde{t} + 0.001)}{\ln 0.001},\nonumber\\ \\
    \frac{r_{s}^{\rm SIDM}(t)}{r_{s,0}^{\rm CDM}} &=& 0.7178 + 0.1026 \tilde{t} + 0.2474 \tilde{t}^2  - 0.4079 \tilde{t}^6 
    + (1 - 0.7178) \frac{\ln (\tilde{t} + 0.001)}{\ln 0.001},\\
    \frac{r_c^{\rm SIDM}(t)}{r_{s,0}^{\rm CDM}} &=& 2.555 \sqrt{\tilde{t}} - 3.632 \tilde{t} + 2.131 \tilde{t}^2 -1.415 \tilde{t}^3 + 0.4683 \tilde{t}^4,
    \label{eq:rc}
\end{eqnarray}
we compute the density profile parameters of the SIDM subhalo at time $t$.

We adopt the formation redshift $z_f$ from ref.~\cite{Correa:2014xma}:
\begin{equation}
    z_f = -0.0064[\log (m_{\rm vir,0}/M_\odot)]^2+0.0237\log(m_{\rm vir,0}/M_\odot)+1.8837,\label{eq:zf}
\end{equation}
where $m_{\rm vir,0}$ is the extrapolated subhalo's virial mass at present $(z=0)$, if there were no tidal processes or SIDM effects.
To connect the virial mass at accretion $m_a$ and at present $m_{\rm vir,0}$, we adopt eq.~(19) of ref.~\cite{Correa:2014xma}.
The formation time $t_f$ is then computed as
\begin{equation}
    t_f = \frac{1}{H_0}\int_{z_f}^\infty \frac{dz}{(1+z)\sqrt{\Omega_m(1+z)^3+\Omega_\Lambda}},
\end{equation}
and the lookback time corresponding to $z_f$ is $t_L(z_f) = t_U - t_f$, where $t_U \approx 13.8$~Gyr is the age of the Universe. {We plan to revisit the assumption of a universal formation time for subhalos of a given mass in future work. For our current study, the version of the parametric model that uses Eq.~(\ref{eq:zf}) is sufficiently accurate, as discussed above.}

Finally, the core-collapse timescale, $t_c$, can be expressed as \cite{Balberg:2002ue,Pollack:2014rja}
\begin{equation}
    t_c = \frac{150}{C (\sigma_{\rm eff}/m_\chi) \rho_s^{\rm CDM}r_s^{\rm CDM}}\frac{1}{\sqrt{4\pi G\rho_s^{\rm CDM}}},
    \label{eq:collapse timescale}
\end{equation}
where $C = 0.75$~\cite{Koda:2011yb,Essig:2018pzq,Nishikawa:2019lsc,Yang:2022zkd}.
This timescale is a function of the cosmic time $t$, through the evolving density profile parameters $\rho_s^{\rm CDM}$ and $r_s^{\rm CDM}$ as well as the effective scattering cross section per SIDM particle mass, $(\sigma_{\rm eff}/m_\chi)$. We introduce this effective cross section and our SIDM models in the following subsection.

To summarise, we use Eqs.~(\ref{eq:V_max})--(\ref{eq:r_c}) to obtain $(V_{\rm max}, r_{\rm max})$ for SIDM from those for CDM, and convert them to $(\rho_s, r_s, r_c)$ for SIDM via Eqs.~(\ref{eq:Vmax0})--(\ref{eq:rc}).  
As the parameters $(\rho_s, r_s)$ of the CDM density profile at any redshifts are entirely determined by $(m_a, z_a, c_a)$, we can also obtain $(\rho_s, r_s, r_c)$ for SIDM in a deterministic manner once we specify $(m_a, z_a, c_a)$.
In other words, we have complete functional forms for any parameters of the subhalo density profile that depends only on $(m_a,z_a, c_a)$: $\bm\theta = \bm\theta(m_a,z_a,c_a)$.

\subsection{SIDM models}

\begin{table}[]
    \centering
    \begin{tabular}{lrr} \hline
    & $\sigma_0/m_\chi ~ [\mathrm{cm^2~g^{-1}}]$ & $w ~ [\mathrm{km~s^{-1}}]$ \\ \hline 
    \textbf{Model~I}     &  {\bf 147.1} & {\bf 24.33} \\
    Model~II & $2.4\times 10^{4}$ & 1 \\
    Model~III & 147.1 & 120 \\
    Model~IV & 5 & 100 \\
    Model~V & 10 & --
    \\ \hline
    \end{tabular}
    \caption{SIDM model parameters. Model~I is adopted as our fiducial model~\cite{Turner:2020vlf,Yang:2022mxl}, and used throughout the work, including to compare SASHIMI-SIDM predictions with cosmological simulation results \cite{Yang:2022mxl}. The other models are discussed in Section~\ref{sec:SIDM models}. Model III was simulated in ref.~\cite{Nadler:2023nrd}, and Model~V is a velocity-independent case chosen for illustrative purposes.}
    \label{tab:SIDM models}
\end{table}

We consider the velocity-dependent scattering cross section
\begin{equation}
    \frac{d\sigma}{d\cos\theta} = \frac{\sigma_0}{2[1+(v/w)^2\sin^2(\theta/2)]^2},
    \label{eq:dsigmadcostheta}
\end{equation}
which is characterised by two parameters: the cross section amplitude $\sigma_0$, and the velocity scale of the transition between velocity-dependent and velocity-independent scattering, $w$.
As default values, we adopt $\sigma_0/m_\chi = 147.1$~cm$^2$~g$^{-1}$ and $w = 24.33$~km~s$^{-1}$~\cite{Turner:2020vlf, Yang:2022mxl} for most of the present paper.

\begin{figure}
    \centering
    \includegraphics[width=12cm]{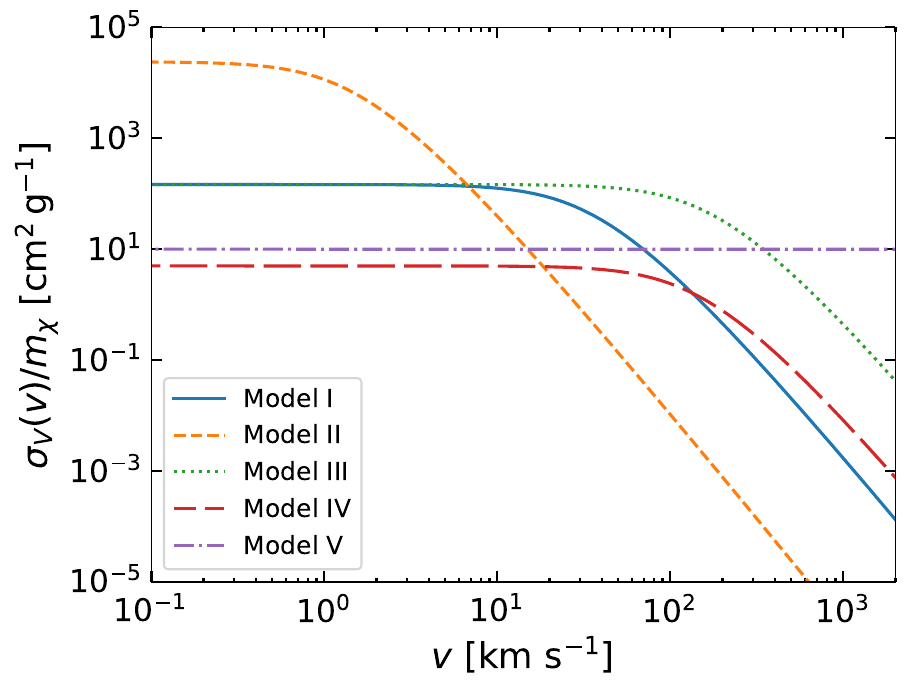}
    \caption{Viscosity cross section per unit dark matter mass, $\sigma_{V}/m_{\chi}$, as a function of relative velocity for five benchmark SIDM models. Our fiducial SIDM model (Model~I) with $\sigma_0/m_\chi = 147.1$~cm$^{2}$~g$^{-1}$ and $w = 24.33$~km~s$^{-1}$ \cite{Turner:2020vlf,Yang:2022mxl} is adopted throughout the paper, while the other models (table~\ref{tab:SIDM models}) are discussed in Section~\ref{sec:SIDM models}. Model III was simulated in ref.~\cite{Nadler:2023nrd}, and Model V is a velocity-independent scenario for illustrative purposes.}
    \label{fig:sigma_viscosity}
\end{figure}

In figure~\ref{fig:sigma_viscosity}, we show the `viscosity' scattering cross section~\cite{Yang:2022hkm} after integrating eq.~(\ref{eq:dsigmadcostheta}) over $\cos\theta$ with the weight of $3\sin^2\theta /2$:
\begin{equation}
    \sigma_V = \frac{6 \sigma_0 w^4}{v^4} \left[ \left(1+\frac{2w^2}{v^2}\right) \ln \left( 1 + \frac{v^2}{w^2} \right) - 2 \right],
    \label{eq:sigma_viscosity}
\end{equation} for this and other models from the previous literature~\cite{Gilman:2021sdr, Yang:2023jwn,Nadler:2023nrd}; we summarise the SIDM parameters of these models in table~\ref{tab:SIDM models}. Along with four velocity-dependent SIDM models (Models I, II, III, and IV), we also consider a velocity-independent model with $\sigma_0/m_{\chi}=10~\mathrm{cm^2~g^{-1}}$ for illustrative purposes only, since observations on galaxy cluster scales imply $\sigma_0/m_{\chi}<0.1~\mathrm{cm^2~g^{-1}}$ for velocity-independent scattering~\cite{Kaplinghat:2015aga,Andrade:2020lqq}.

The viscosity cross section is suggested as a better representation of an angle-integrated quantity than the total cross section $\sigma_{\rm total} = \sigma_0/(1+v^2/w^2)^2$~\cite{Yang:2022hkm}.
The effective cross section $\sigma_{\rm eff}$ is obtained by~\cite{Yang:2022hkm,Yang:2022zkd}
\begin{equation}
    \sigma_{\rm eff} = \frac{1}{512v_{\rm eff}^8}\int_0^\infty dv\int_{-1}^1 d\cos\theta \frac{d\sigma}{d\cos\theta} v^7 \sin^2\theta\exp\left[-\frac{v^2}{4v_{\rm eff}^2}\right],
\end{equation}
which depends on subhalo density profile via $v_{\rm eff} = 0.64V_{\rm max}^{\rm CDM}$~\cite{Outmezguine:2022bhq}.

\subsection{Validity of SASHIMI in sub-resolution regimes}

One of the advantages of adopting semi-analytical models like SASHIMI over approaches based on numerical simulations is that they are free from issues related to numerical resolution in cosmological simulations.
If the physical ingredients such as our ePS formalism and parametric SIDM model are assumed to be valid, then our SASHIMI-SIDM predictions will be trustworthy.
We may therefore make predictions for arbitrarily small subhalos under this assumption. Before doing so, we will demonstrate in Section~\ref{sec:Results} that SAHSIMI-SIDM matches the predictions of cosmological SIDM simulations in resolved regimes.

The ePS formalism has been tested and proven to be a valid tool across a wide range of masses, redshifts, and environments (e.g.,~\cite{Zheng:2023myp}).
The parametric model for SIDM discussed in Section~\ref{sec:Parametric model of SIDM} depends on the cosmic time since the subhalos' formation, normalized by the collapse timescale.
The collapse timescale depends on the scattering cross section (per unit dark matter mass) and halo density profile parameters, and thus our ansatz is that this dependence continues to hold at much smaller scales.
This, however, must be tested against future numerical simulations with higher resolution.

Below, we provide the first direct comparison between SASHIMI-SIDM predictions and high-resolution zoom-in cosmological simulations of SIDM.
In order to do so and also to make predictions for various observables, we will make appropriate cuts in the following sections.

\section{Model predictions and validations}
\label{sec:Results}

Using SASHIMI, we generate a list of subhalos in a Milky Way-mass host with $M_{200} = 10^{12} M_\odot$ at $z=0$.
The parameters related to the subhalo's density profile are mass at both accretion and at present ($z=0$), redshift at accretion $z_a$, $\rho_s$ and $r_s$ for both CDM and SIDM, the core radius $r_c$ for SIDM, and the tidal truncation radius $r_t$.

The properties of each subhalo are entirely characterised by three parameters $m_a$, $z_a$, and $c_{a}$, throughout its evolution history.
We generate these quantities following their distributions; i.e., the ePS $d^2N_{\rm sh}/(dm_adz_a)$ for the former two, and the log-normal distribution $P(c_a)$ for the latter.
We assign each subhalo entry in the list a weight factor, i.e., the ensemble average of the number of such a subhalo: $w_i = \langle N_{{\rm sh},i}\rangle$.
This approximates the distribution of $(m_a, z_a, c_a)$ as 
\begin{equation}
    \int dz_a \int dm_a \frac{d^2N_{\rm sh}}{dm_adz_a}\int dc_a P(c_a|\bar c_{\rm vir}(m_a,z_a),\sigma_{\log\sigma}) \longrightarrow \sum_i w_i.
\end{equation}
We then generate the distribution function of any combination of subhalo parameters, ${\bm \theta}$,
\begin{equation}
    \frac{dN_{\rm sh}}{d\bm\theta} = \int dz_a \int dm_{a}\frac{d^2N_{\rm sh}}{dm_{a}dz_a}\int dc_{a} P(c_{a}|\bar c_{\rm vir}(m_{a},z_a), \sigma_{\log c}) 
    \delta_D\left(\bm\theta-\bm\theta(m_{a},z_a,c_{a})\right),
    \label{eq:dNdtheta}
\end{equation}
by making a (multi-dimensional) histogram with weights $w_i$.

{Note in the current version of SASHIMI, we model the average properties of the subhalo population within the host's virial volume, without including spatial or orbital information directly. This means that the effects of subhalo orbits on tidal mass loss and gravothermal collapse are not captured. Previous studies, such as refs.~\cite{Zeng:2021ldo, Andrade:2023fgr}, highlight the importance of orbital parameters in shaping the evolution of collapsed and cored subhalos, particularly because core collapse can be accelerated by tidal stripping~\cite{Nishikawa:2019lsc,Sameie:2019zfo}. Incorporating these orbital distributions would allow us to account for the radial dependence of tidal interactions and collapse timescales. Additionally, we do not account for the evaporation effect between host and subhalo dark matter particles, which becomes significant when the cross section at the host mass scale exceeds $1~\mathrm{cm}^2~\mathrm{g^{-1}}$~\cite{Nadler:2020ulu}. This effect can be included in the parametric model; see ref.~\cite{Yang:2024uqb} for details. We plan to extend SASHIMI in future work to incorporate these effects, improving the fidelity of our predictions for the core-collapsed and cored subhalo populations.}

\subsection{Distribution of density profile parameters}
\label{sec:Distribution of density profile parameters}

\begin{figure}
    \centering
    \includegraphics[width=7.5cm]{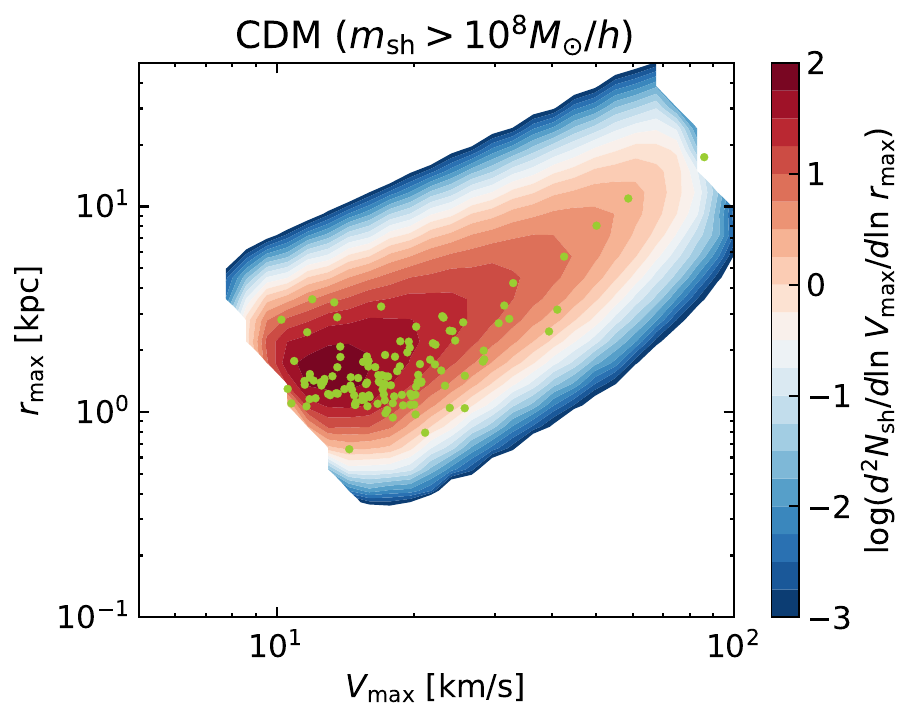}
    \includegraphics[width=7.5cm]{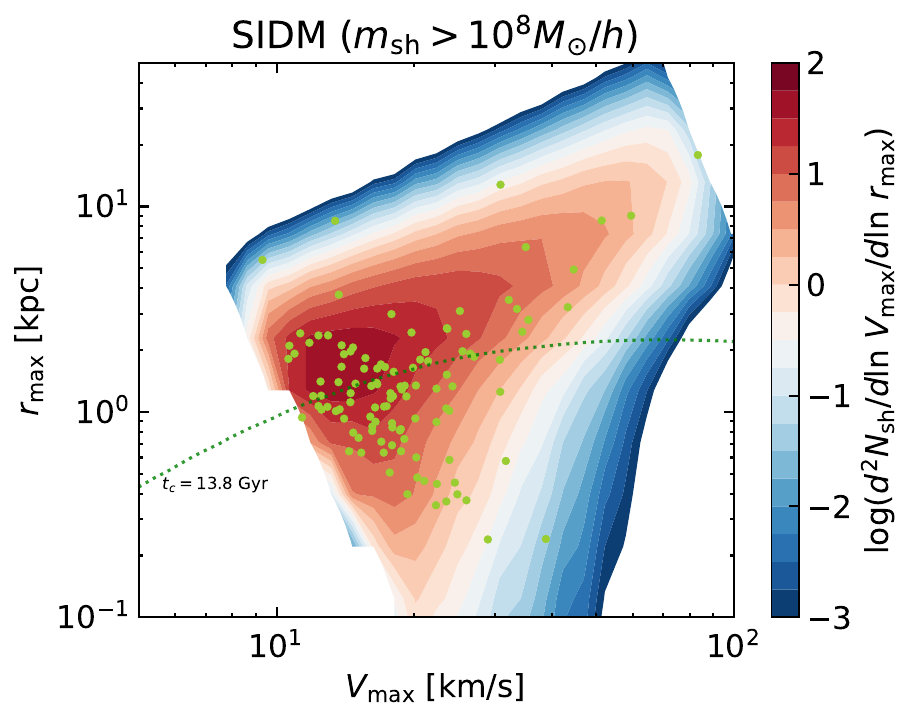}
    \includegraphics[width=7.5cm]{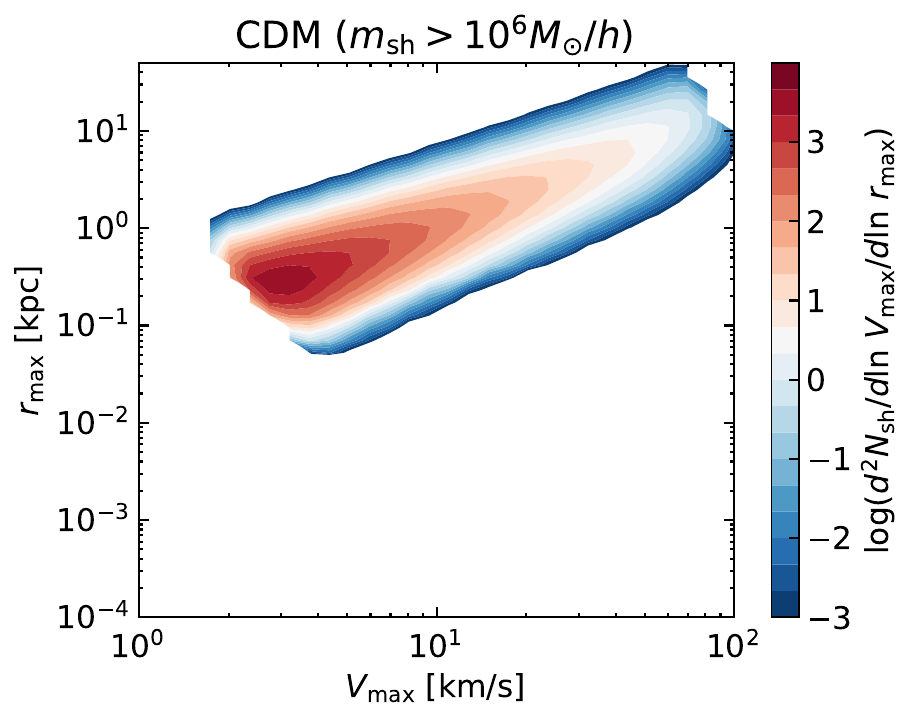}
    \includegraphics[width=7.5cm]{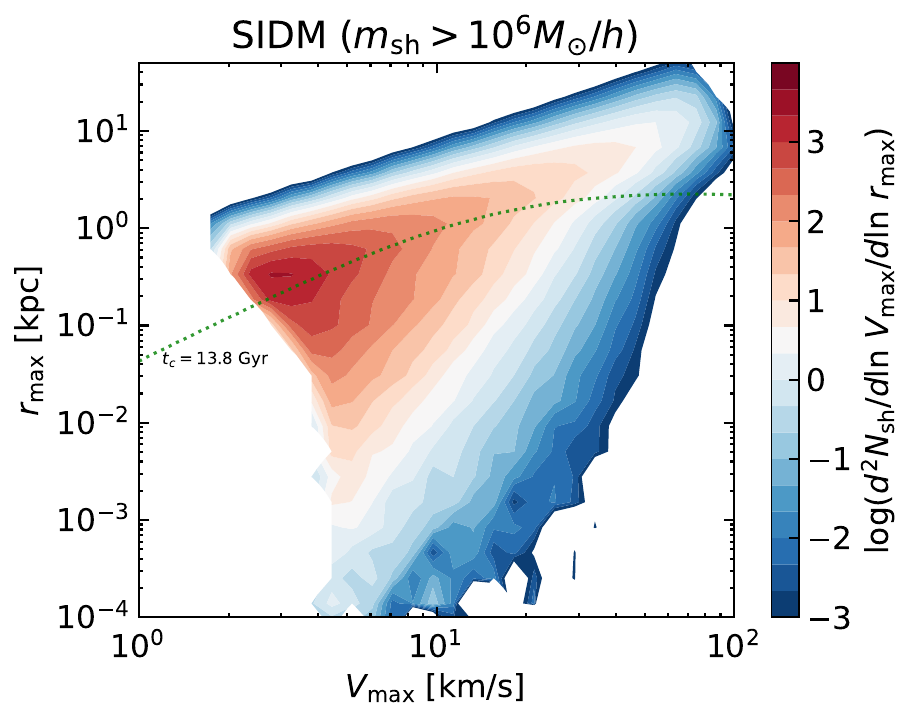}
    \caption{Subhalo number distribution in the $(V_{\rm max}, r_{\rm max})$ parameter space for CDM ({\it left}) and SIDM ({\it right}) for the SIDM Model~I (table~\ref{tab:SIDM models}). The subhalos shown have masses larger than $10^{8}M_\odot/h$ ({\it top}) and $10^{6}M_\odot/h$ ({\it bottom}). Green data points in the top panels are from cosmological zoom-in simulations of SIDM Model I~\cite{Yang:2023jwn}. The dotted curve in each panel shows a contour of the collapse timescale, obtained by setting $t_c$ from eq.~(\ref{eq:collapse timescale}) equal to the age of the Universe, 13.8~Gyr. The SIDM subhalos below this curve are expected to be in core-collapse phase at $z=0$.
    }
\label{fig:contour_Vmax_rmax}
\end{figure}

Figure~\ref{fig:contour_Vmax_rmax} shows the density contour in the two-dimensional parameter space: $\bm\theta = (V_{\rm max}, r_{\rm max})$, computed using eq.~(\ref{eq:dNdtheta}).
The top panels are for the subhalos with the masses within the truncation radius at $z=0$ that are larger than $10^8M_\odot/h$.
We compare the distributions in the case of SIDM with that of CDM.
We also plot results from cosmological zoom-in simulations of SIDM Model I, presented in the previous study by some of the present authors~\cite{Yang:2023jwn}, as green data points. Our SASHIMI-SIDM predictions broadly agree with these simulation results for both the CDM and SIDM cases. We leave a detailed statistical analysis based on matching individual subhalos in cosmological simulations to future work.

In the bottom panels of figure~\ref{fig:contour_Vmax_rmax}, we extend the SASHIMI-SIDM predictions to the sub-resolution regime with the subhalo mass threshold of $10^6 M_\odot/h$.
The resulting $r_{\rm max}$ distribution extends to much lower values compared to the top panel, while the $V_{\rm max}$ distribution hardly changes.
This follows because the rotation curves of collapsed SIDM subhalos reach a similar maximum velocity at a much smaller radius, reflecting their higher concentration.

\begin{figure}
    \centering
    \includegraphics[width=7.5cm]{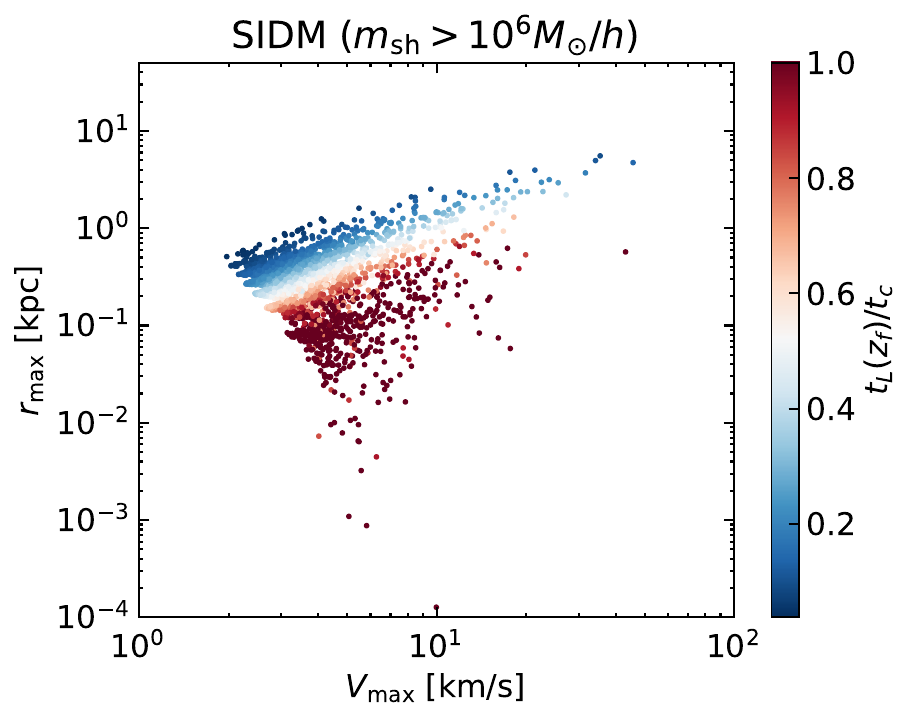}
    \includegraphics[width=7.5cm]{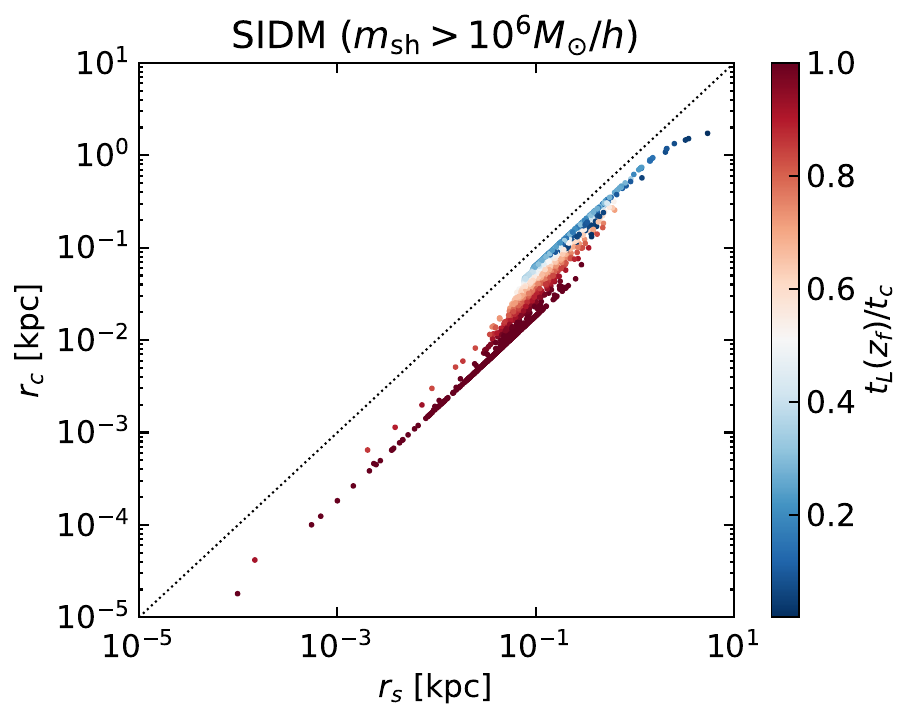}
    \caption{A realisation of a SASHIMI-SIDM subhalo population with masses heavier than $10^{6} M_\odot/h$, on the $(V_{\rm max}, r_{\rm max})$ ({\it left}) and $(r_s, r_c)$ ({\it right}) planes for the SIDM Model~I. The diagonal dotted line in the right panel show the one-to-one $r_c = r_s$ relation. The colour corresponds to the lookback time to the subhalo formation redshift, $t_L(z_f)$, in units of the collapse timescale $t_c$. The colours are saturated for all the subhalos in the core-collapse phase, $t_L(z_f)/t_c>1$, as the evolution has been stopped once it reaches $t_L(z_f)/t_c = 1.1$.}
    \label{fig:scatterSIDM}
\end{figure}

Figure~\ref{fig:scatterSIDM} shows a subhalo population generated by Monte-Carlo sampling the SASHIMI-SIDM prediction with $\bm\theta = (V_{\rm max}, r_{\rm max})$ and $(r_s, r_c)$, in the left and right panels, respectively.
The colour of these scattered points shows the lookback time to the subhalo formation redshift $t_L(z_f)$ in units of the collapse timescale $t_c$.
The subhalos in the collapsing phase, $t_L(z_f)/t_c \approx 1$, show substantially smaller $r_{\rm max}$ values for the same $V_{\rm max}$ values.
The core radius $r_c$ is strongly correlated with the scale radius $r_s$. 
The relation saturates around $r_s \approx 5 r_c$ for the subhalos in the core-collapse phase.

\subsection{Mass function and fraction of collapsed subhalos}

\begin{figure}
    \centering
    \includegraphics[width=7.5cm]{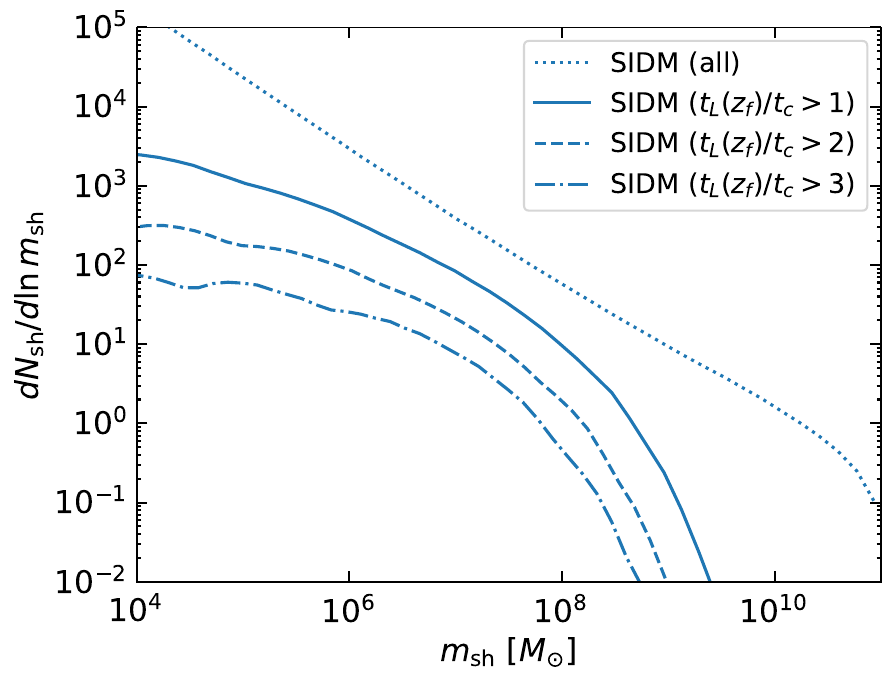}
    \includegraphics[width=7.5cm]{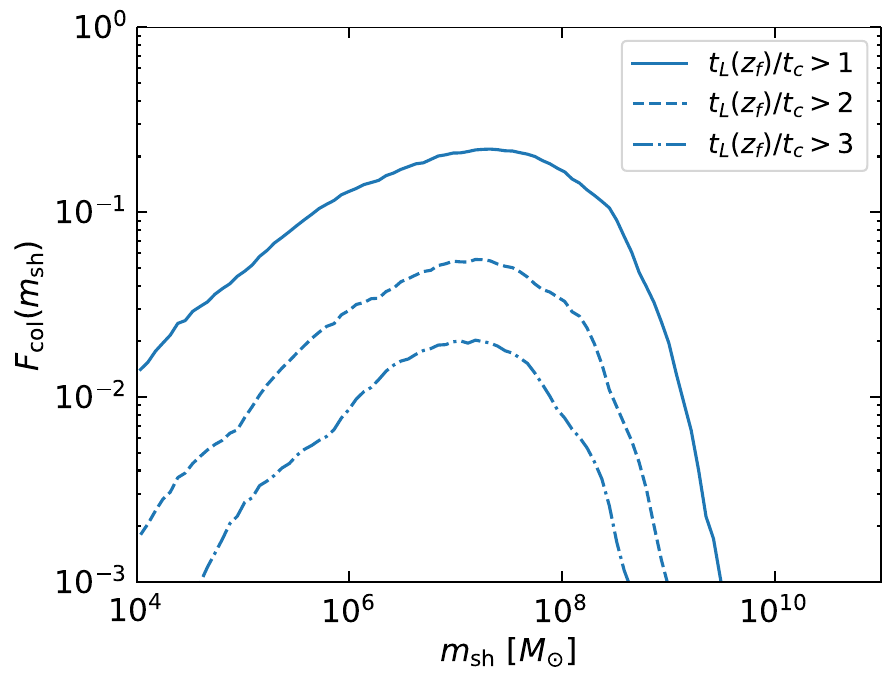}
    \caption{{\it Left}: Subhalo mass function for the SIDM Model~I, which is nearly identical to CDM. The mass function of the collapsed subhalos with $t_L(z_f)/t_c > 1$, 2, and 3 are also shown.
    {\it Right}: Collapse fraction $F_{\rm col}$: i.e., the ratio of mass functions of collapsed subhalos compared to the total population.}
    \label{fig:SHMF}
\end{figure}

We discuss the subhalo mass function, for which $\bm \theta=m$ is the present-day bound subhalo mass.
In figure~\ref{fig:SHMF} (left), we show $dN_{\rm sh}/dm$ for the subhalos.
We also show the mass function of the subhalos that have already collapsed with $t_L(z_f)/t_c >1$, 2, and 3.
The fraction of collapsed subhalos is shown in the right panel. The subhalo mass function is hardly changed compared to that of the CDM case.
This is based on the assumption that the ePS formalism does not change, and that the subhalo masses are not significantly affected by self-interactions (e.g.,~\cite{Yang:2022mxl}).
Strikingly, the fraction of core-collapsed subhalos with $t_L(z_f)/t_c>1$ peaks at $\sim$30\% for the subhalos with the masses of $10^7$--$10^8 M_\odot$.

We can understand this result analytically as follows. The mass scale where the core-collapsed fraction peaks is determined by the underlying SIDM cross section and the mass--concentration relation. We parameterize the velocity dependence of the effective cross section as $\sigma_{\rm eff}\propto\sigma_0 v^{-n}$, where the power-law exponent $n$ is a number varying from $0$ to $4$ in our SIDM scenarios. For a given halo, $v\sim V_{\rm max}\propto(\rho^{\rm CDM}_s)^{1/2}r^{\rm CDM}_s$. Hence, from eq.~(\ref{eq:collapse timescale}), we see that the core-collapse timescale scales as 
\begin{equation}
    t_c\propto \frac{1}{(\sigma_{\rm eff}/m_\chi)r^{\rm CDM}_s(\rho^{\rm CDM}_s)^{3/2}}  
      \propto (\sigma_0/m_\chi)^{-1}m^{(n-1)/3}c^{(n-7)/2}, \label{eq:cc_scaling}
\end{equation}
where we have used the scaling relations $\rho^{\rm CDM}_s\propto c^{3}$ and $r^{\rm CDM}_s\propto m^{1/3}/c$. In the limit $n=0$, $t_c\propto(\sigma/m_\chi)^{-1}m^{-1/3}c^{-7/2}$~\cite{Essig:2018pzq,Nadler:2023nrd}.

\begin{figure}
    \centering
    \includegraphics[width=7.5cm]{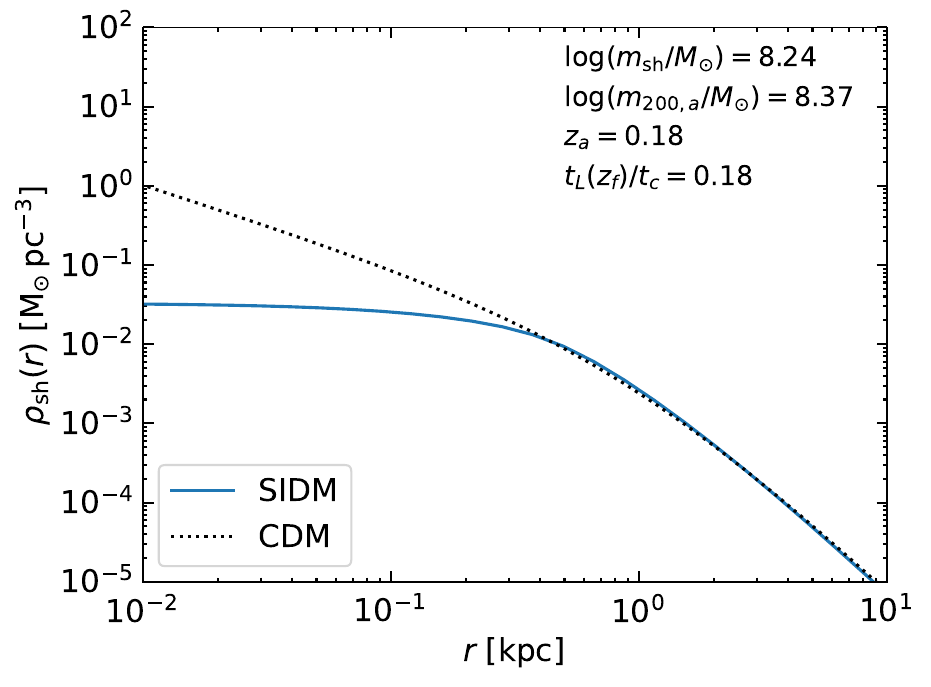}
    \includegraphics[width=7.5cm]{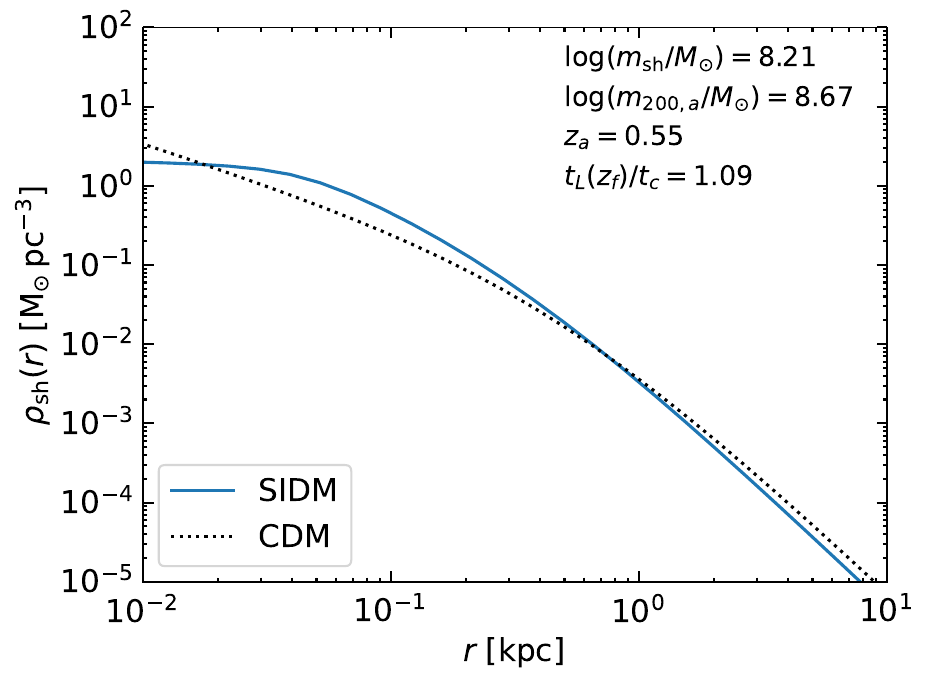}
    \includegraphics[width=7.5cm]{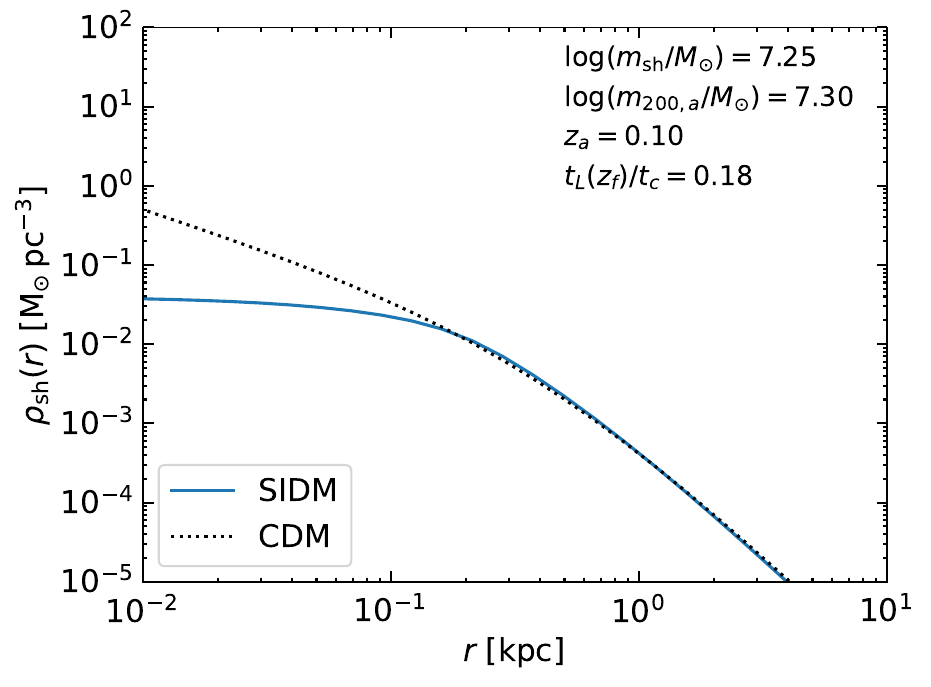}
    \includegraphics[width=7.5cm]{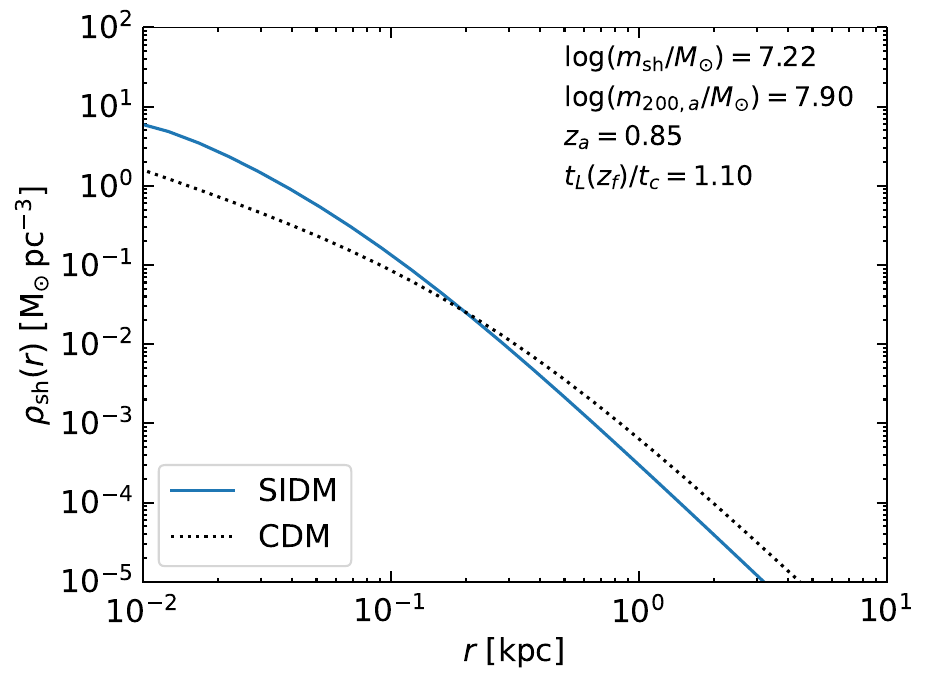}
    \caption{Density profiles of both CDM and SIDM subhalos with the parameters shown: the present bound mass $m_{\rm sh}$, mass at accretion $m_{200, a}$, redshift at accretion $z_a$, and the lookback time of the subhalo formation $t_L(z_f)$ in units of collapse timescale $t_c$. The left panels show subhalos in a core-forming phase ($t_L(z_f)/t_c\sim 0.2$), and the right panels show those in a collapsing phase ($t_L(z_f)/t_c \gtrsim 1$). 
    }
    \label{fig:density}
\end{figure}

According to our assumed mass--concentration relation from ref.~\cite{Correa:2015dva}, $c\propto m^{-0.036}$ in the low-mass regime. Thus, for halos that probe the velocity-independent part of the SIDM cross section ($n=0$), $t_c\propto m^{-0.2}$, implying that the core-collapsed fraction decreases toward lower halo masses. Meanwhile, for halos in the velocity-dependent regime of the cross section ($n=4$), $t_c\propto m^{1.1}$ and higher-mass halos collapse more slowly. {This behaviour, including the minimum of $t_c$, was previously identified by combining cosmological CDM simulations and semi-analytical models~\cite{Shah:2023qcw}. Our semi-analytical approach, however, allows us to explicitly quantify the resulting core-collapsed fraction across a broad mass spectrum, illustrating the peak and its dependence on the SIDM parameters. Crucially, our predictions span subhalo masses below the resolution limit of any current cosmological SIDM simulation.} In the context of our SIDM parameterisation, the amplitude of the core-collapse fraction peak is set by $\sigma_0$, and the mass scale of the peak is set by $w$. For SIDM Model I, the core-collapse fraction peaks at~$\approx 30\%$, consistent with cosmological zoom-in simulation results for this model~\cite{Yang:2022mxl}. This peak occurs for $\approx 10^8~M_{\mathrm{\odot}}$ subhalos, which have maximum circular velocities comparable to the $w=24.33~\mathrm{km\ s}^{-1}$ in this model. In Section~\ref{sec:SIDM models}, we present core collapse fractions for other SIDM models to expand on these predictions and confirm the dependence of the core-collapse fraction on $\sigma_0$ and $w$.

\subsection{Density profiles}
\label{sec:Density profiles}

We demonstrate several examples of subhalos' density profiles, shown in figure~\ref{fig:density}.
These subhalos were randomly sampled following the subhalo distribution generated with SASHIMI-SIDM under conditions of $t_L(z_f)/t_c\approx 0.18$ (left panels) and $t_L(z_f)/t_c\approx 1.1$ (right panels).
We chose subhalos, whose masses fall within a $10^7$--$10^9 M_\odot$ range, for the core-forming and core-collapsed cases.

As expected, if the core-collapse timescale $t_c$ is shorter than the lookback time corresponding to the halo formation $t_L(z_f)$, the subhalo is in the core-collapse phase.
To make our predictions conservative (i.e., such that we underestimate the density of the collapsed subhalos), we stop our calculations and freeze the density profile evolution according to the SIDM effects once the subhalo reaches $t_L(z_f)/t_c = 1.1$.
We note that the subhalos will keep evolving after reaching the collapsing time $t_c$ and might form a black hole at the subhalo centre in extreme scenarios (e.g.,~\cite{Feng:2020kxv}).

\section{Applications}
\label{sec:Discussion}

We apply the SASHIMI-SIDM results to predict a few observables to probe the particle properties of dark matter using future experiments and observations.
More detailed analyses are slated for several follow-up studies. 
Here, we demonstrate a couple of simple examples of applications.

\subsection{Distinguishing SIDM models}
\label{sec:SIDM models}

First, we discuss the dependence on the SIDM parameters $\sigma_0/m_\chi$ and $w$.
The left panel of figure~\ref{fig:Bsh} shows the fraction of collapsed subhalos with $t_L(z_f)/t_c > 1$ as a function of the subhalo mass.
The parameters and cross sections of these SIDM models are as shown in table~\ref{tab:SIDM models} and figure~\ref{fig:sigma_viscosity}, respectively.
As expected, the larger the cross section $\sigma_0$ is, so is the collapse fraction.
The velocity scale $w$, which sets the transition scale from $\sigma_0 \propto v^0$ to $\propto v^{-4}$, decides the mass scales of the core-collapsed subhalos.
Particularly for SIDM Model~II, which saturates at the large value of $\sigma_0/m_\chi = 2.4\times 10^4$~cm$^2$~g$^{-1}$, nearly all the subhalos with the masses of 1--$10^6 M_\odot$ collapse. 

We emphasize that, in the subhalo mass regime where simulations can resolve core collapse, these results are consistent with previous studies~\cite{Yang:2022mxl,Nadler:2023nrd,Shah:2023qcw}. However, our semi-analytical approach allows us to predict the core-collapse fraction down to very low subhalo masses and to reveal its turnover. It will be interesting to design cosmological simulations that capture this turnover in future work.
Future surveys of Milky Way dwarf galaxies and follow-up measurements of their dark matter density profiles will help determine the population of dwarfs that are potentially in the collapse phase.
They may further help distinguish different SIDM models, where our method provides essential tools.

\subsection{Subhalo boost of dark matter annihilation}
\label{sec:annihilation boost}

Next, we discuss a possibility that the SIDM particles self-annihilate into Standard Model final states (motivated by several particle physics models, e.g.,~\cite{Tulin:2013teo, Wu:2022wzw}).
Since the rate of self-annihilation depends on the density squared, the subhalos in the core-collapse phase will enhance the annihilation signal, and this signature will depend on the SIDM model parameters through the core-collapsed fraction.
We estimate the total luminosity of Standard Model particles such as photons from the dark matter annihilation in all subhalos within the host's virial radius by
\begin{equation}
    L_{\rm sh,tot} \propto \sum_i w_i \int dr 4\pi r^2 \rho_{{\rm sh},i}^2(r).
\end{equation}
By dividing this quantity by the luminosity of the main halo component by assuming the NFW profile (i.e., CDM without subhalos), we define the annihilation boost factor: $B_{\rm sh} \equiv L_{\rm sh, tot}/ L_{\rm host, NFW}^0$.
We then compare this quantity for CDM and SIDM, and investigate the dependence on the subhalo masses considered. Our assumption that the main halo is CDM-like is reasonable even for our SIDM models because baryons contract the inner host halo~\cite{Kaplinghat:2013xca,Sameie:2018chj,Robles:2019mfq,Rose:2022mqj}. This choice allows us to isolate the effects of SIDM on subhalo populations.
We note that the effect of sub-subhalos and finer structures is not included in the following calculations, and hence the results in this subsection is conservative.

\begin{figure}
    \centering
    \includegraphics[width=7.5cm]{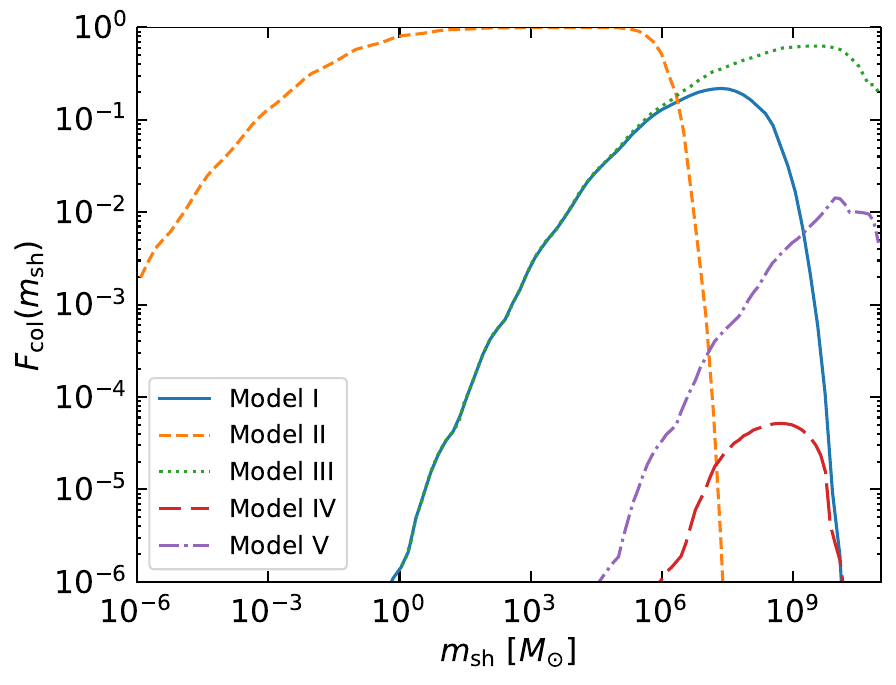}
    \includegraphics[width=7.5cm]{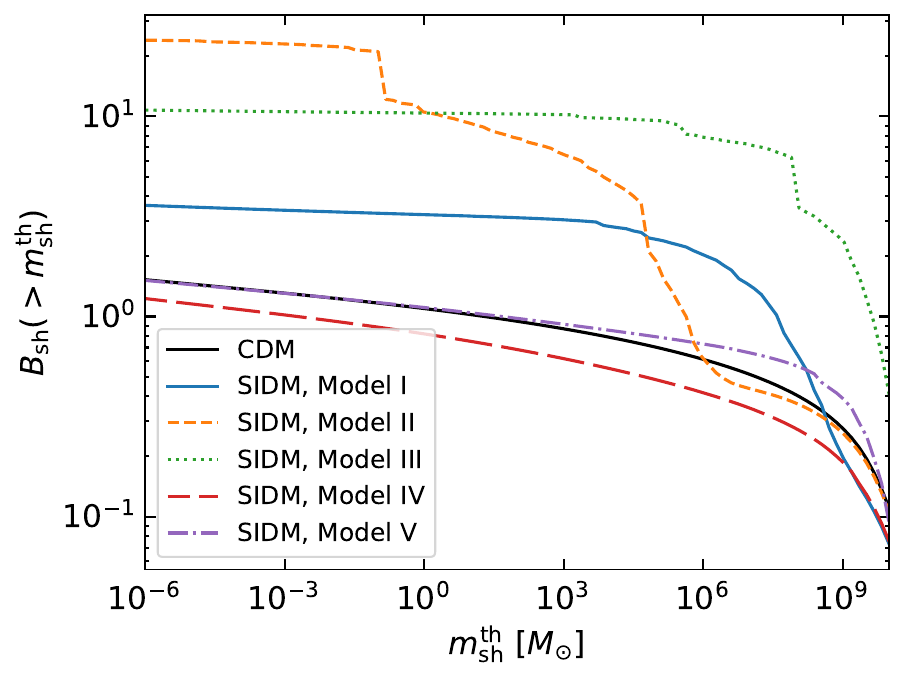}
    \caption{\textit{Left}: Core-collapsed fractions (same as figure~\ref{fig:SHMF} right panel, $t_L(z_f)/t_c > 1$), for various SIDM models. \textit{Right}: The annihilation boost factor---the ratio of the luminosity from dark matter annihilation in all subhalos throughout the virial radius, relative to the smooth CDM host halo---as a function of threshold subhalo mass $m_{\rm sh}^{\rm th}$. For the parameters of SIDM models, see table~\ref{tab:SIDM models}.}
    \label{fig:Bsh}
\end{figure}

The right panel of figure~\ref{fig:Bsh} shows the annihilation boost factor due to the subhalos more massive than a threshold mass $m_{\rm sh}^{\rm th}$, $B_{\rm sh}(>m_{\rm sh}^{\rm th})$. 
First, we focus on comparing SIDM Model~I to CDM.
As we include smaller subhalos, the CDM boost factor gradually increases and exceeds 1 for $m_{\rm sh} > M_\odot$~\cite{Hiroshima:2018kfv, Ando:2019xlm}.
On the other hand, for the SIDM model, the contribution from subhalos with masses larger than $\sim 10^{8}M_\odot$ is small as many of them are cored.
A clear transition happens once the subhalos with masses smaller than $\sim$10$^8M_\odot$ are included.
Because of the large population of core-collapsed subhalos ($t_L(z_f)/t_c>1$) that have much steeper profile than the CDM case, the signals of dark matter annihilation are boosted up to a factor of 4.
Given the null observations of dark matter annihilation in any existing gamma-ray data, the model will be stringently constrained by using the existing gamma-ray data from the Milky Way halo, dwarf spheroidal galaxies~\cite{Ando:2020yyk, Horigome:2022gge}, unassociated Fermi sources, clusters of galaxies, and unresolved gamma-ray background.
More careful analyses of these targets are the subject for our future study. For example, it will be interesting to model the spatial and orbital dependence of subhalo mass-loss rates, rather than calculating orbit-averaged quantities, in order to predict the radial dependence of the annihilation boost.

SIDM models that predict larger fraction of collapsed subhalos yield more conspicuous dark matter annihilation boosts.
In particular, for SIDM Model II, the boost factor reaches $B_{\rm sh}\sim 10$ but only if the halos with masses smaller than $1~M_\odot$ exist.
For SIDM Model III, the larger subhalos with $m_{\rm sh} > 10^9 M_\odot$ could collapse within the Hubble time, and thus enhance the boost factor substantially.

\subsection{Probability distribution of dark matter density}

What is the dark matter density at a given point in the Galaxy?
This is one of the most relevant questions for dark matter direct detection experiments.
It may also have some relevance for several gravitational probes, such as gravitational lensing, gaps in stellar streams, and pulsar timing arrays.
To this end, we predict the probability density function $P(\rho) = dP/d\rho$ of dark matter density at a given radius (e.g., $r = 8$~kpc for the solar system), $\rho(r) = \rho_{\rm sh}(r) + \rho_{\rm h}(r)$, where $\rho_{\rm sh}$ and $\rho_{\rm h}$ are the subhalo and host halo components, respectively.
This is motivated by several earlier studies~\cite{Kamionkowski:2008vw, Kamionkowski:2010mi, Ibarra:2019jac}.

Suppose we are in a dark matter subhalo $i$, which subtends to its truncation radius $r_{t,i}$.
Then, the conditional probability of finding a subhalo density between $\rho_{\rm sh}$ and $\rho_{\rm sh} + d\rho_{\rm sh}$, $P(\rho_{\rm sh}|i)d\rho_{\rm sh}$ is obtained via
\begin{equation}
    P(\rho_{\rm sh}|i)d\rho_{\rm sh} = \left|\frac{dP_{{\rm sh},i}}{dr}\right|\left|\frac{dr}{d\rho_{{\rm sh},i}}\right| \propto \frac{r^2}{|d\rho_{{\rm sh},i}/dr|},
    \label{eq:conditional subhalo PDF}
\end{equation}
where $P_{{\rm sh},i}$ after the first equality is the probability of finding the point of interest  at a radius $r$ from the subhalo centre, which goes as $dP_{{\rm sh},i}/dr \propto r^2$.
The denominator of the last expression $|d\rho_{{\rm sh},i}/dr|$ can be evaluated using the density profile expressions for either the CDM or SIDM case, in eqs.~(\ref{eq:NFW}) and (\ref{eq:density profile SIDM}), respectively.

The volume of subhalo $i$ is $V_{{\rm sh},i} = 4\pi r_{t,i}^3/3$.
We assume that the subhalos' spatial distribution follows the density function $n_{\rm sh}(r) \propto \exp[(2/\alpha)(r/r_{-2})^\alpha]$ independent of the mass, where $\alpha=0.678$, $r_{-2} = 0.81 r_{200}$, and $r_{200}$ is the radius within which the average density is 200 times the critical density~\cite{Springel:2008cc}.
With this assumption, the probability per volume that a random subhalo is located at the radius $r$ from the halo centre, $P_{{\rm sh},V}$, is proportional to $n_{\rm sh}(r)$.
Therefore, at the radius $r$, the probability of finding a subhalo $i$ at our location is $P(i) = w_i V_{{\rm sh},i} P_{{\rm sh},V}$.
Note that the weight factor $w_i$ is needed as $i$ represents a collection of the subhalos specified with the same parameters, whose total number is $w_i$.
Combined with the conditional probability from eq.~(\ref{eq:conditional subhalo PDF}), the total probability density function for $\rho_{\rm sh}$ is
\begin{equation}
    P_{\rm sh}(\rho_{\rm sh})d\rho_{\rm sh} = \sum_i P(i)P(\rho_{\rm sh}|i)d\rho_{\rm sh}.
    \label{eq:P_sh}
\end{equation}

Our model predicts that the volume filling factor, $f_{V,{\rm sh}} \equiv \sum_i P(i)$, is 29\% ($r = 1$~kpc), 22\% (8~kpc), 8.7\% (50~kpc), 4.0\% (100~kpc), and 1.2\% (200~kpc), and these figures hardly change between CDM and SIDM.
The remaining fraction, $1-f_{V,{\rm sh}}$, is in the smooth host-halo component.
Following ref.~\cite{Kamionkowski:2010mi}, we model this host component as
\begin{equation}
    P_{\rm h}(\rho)d\rho = \frac{1-f_{V,{\rm sh}}}{\sqrt{2\pi}\Delta}\exp\left\{-\frac{1}{2\Delta^2}\left[\ln\left(\frac{\rho}{\rho_0(r)}e^{\Delta^2/2}\right)\right]^2\right\}\frac{d\rho}{\rho},
    \label{eq:P_rho_h}
\end{equation}
where $\Delta = 0.2$.
For $\rho_0(r)$, we adopt the NFW profile with $\rho_s = 0.245$~GeV~cm$^{-3}$, $r_s = 23.5$~kpc, which gives $\rho_0(8~{\rm kpc}) = 0.4$~GeV~cm$^{-3}$ for the local dark matter density around the solar system.
Combining with the subhalo component from eq.~(\ref{eq:P_sh}), we have the total probability distribution function
\begin{equation}
    P(\rho)d\rho = P_{\rm h}(\rho)d\rho+P_{\rm sh}(\rho-\rho_{\rm h})d\rho.
\end{equation}
We note that for obtaining the second term, we marginalise over the $\rho_{\rm h}$ following the probability density function, eq.~(\ref{eq:P_rho_h}).

\begin{figure}
    \centering
    \includegraphics[width=7.5cm]{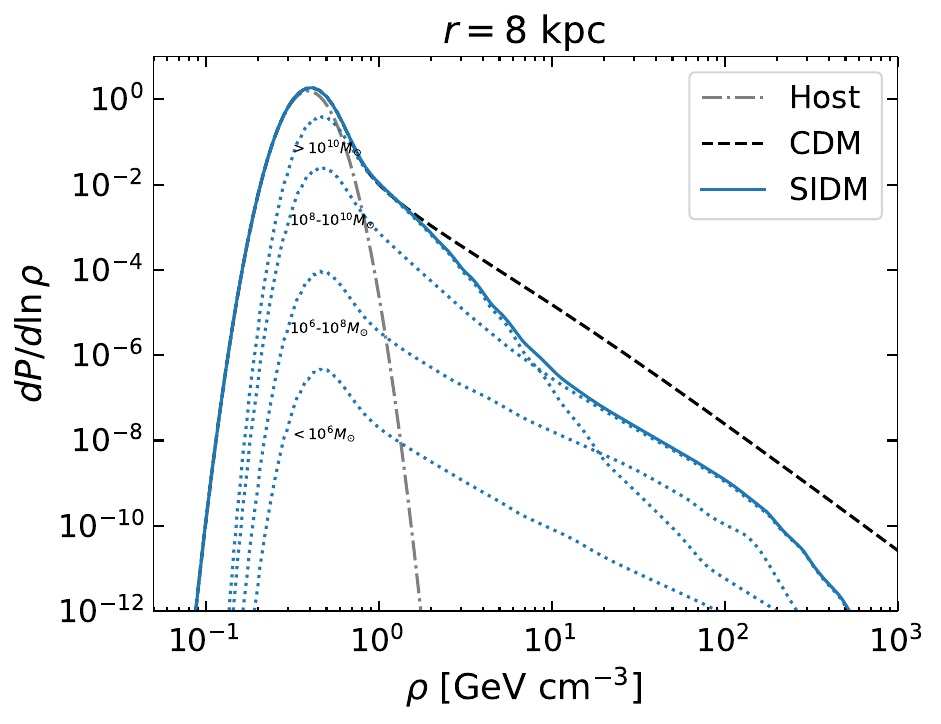}
    \includegraphics[width=7.5cm]{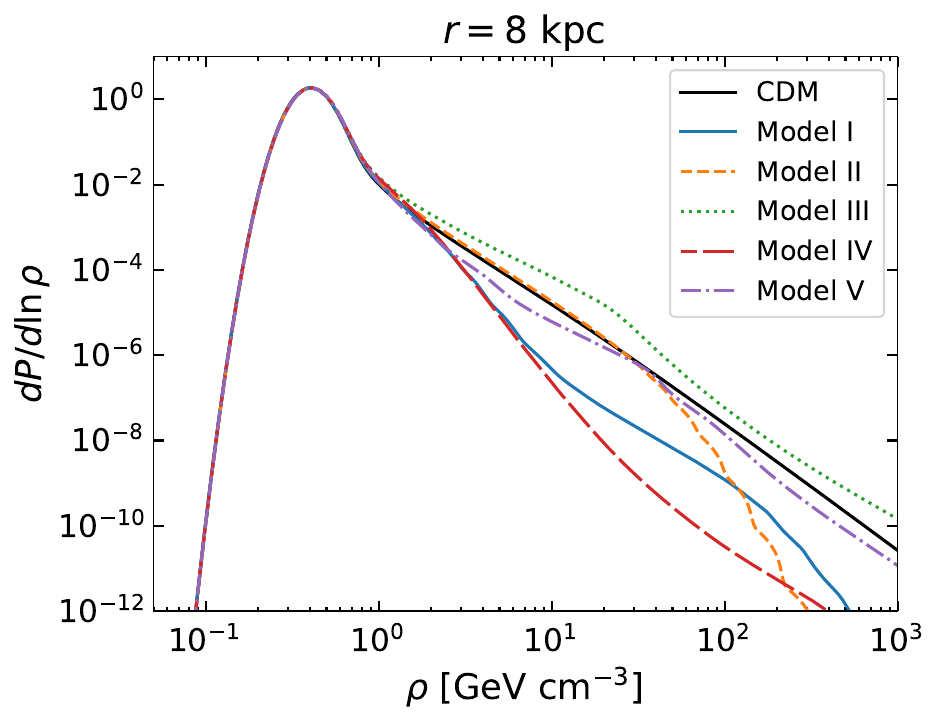}
    \includegraphics[width=7.5cm]{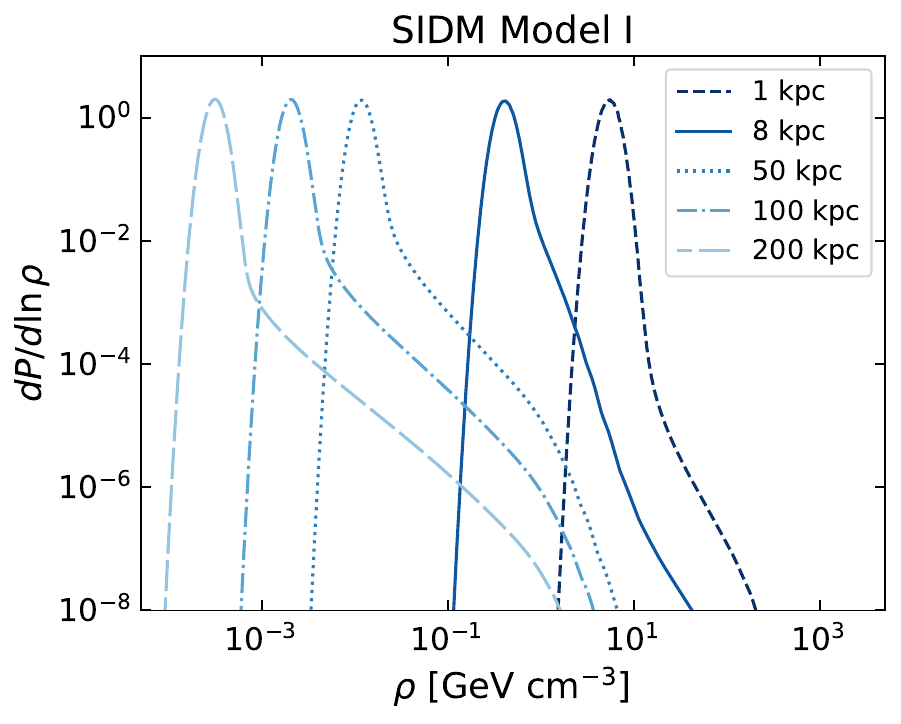}
    \includegraphics[width=7.5cm]{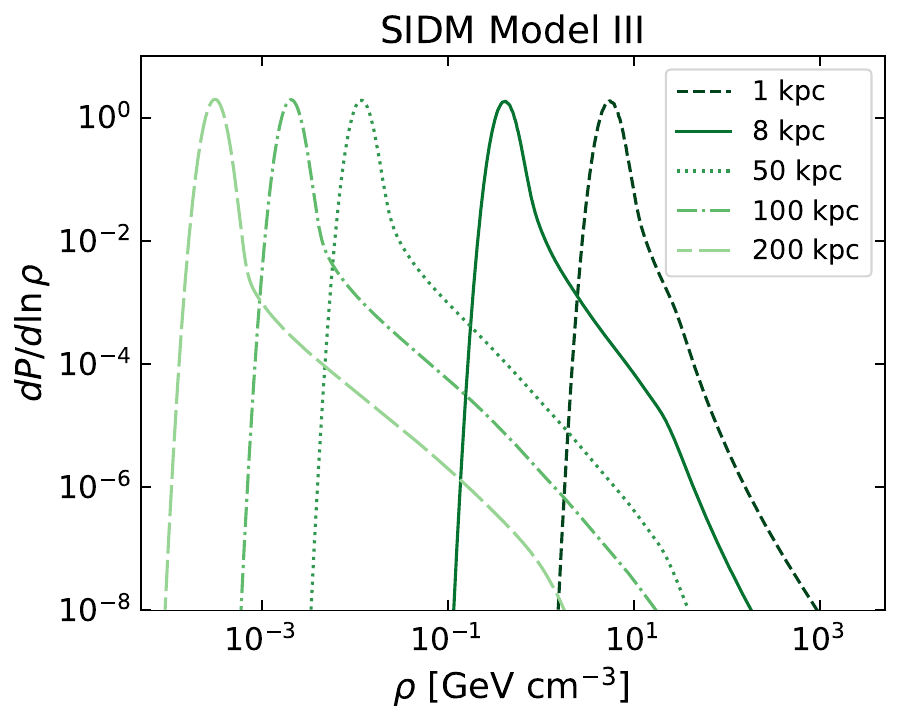}
    \caption{\textit{Top left}: Probability distribution function of the local dark matter density $P(\rho)$ at 8~kpc from the host Milky-Way halo centre (dot-dashed), CDM subhalos and the halo (dashed) and SIDM (Model~I; table~\ref{tab:SIDM models}) subhalos and the halo (solid). 
    The dotted lines show contributions from the SIDM subhalos with masses larger than $10^{10}M_\odot$, between $10^{10}M_\odot$ and $10^{8}M_\odot$, between $10^8M_\odot$ and $10^{6}M_\odot$, and smaller than $10^6M_\odot$, from top to bottom at the left, respectively.  \textit{Top right}: The probability density function at 8~kpc (subhalos and the halo) for SIDM Models~I--V (table~\ref{tab:SIDM models}) and CDM. \textit{Bottom}: The probability density function of SIDM Model~I (\textit{left}) and Model~III (\textit{right}) at $r = 1$, 8, 50, 100, and 200~kpc.}
    \label{fig:PDF_density}
\end{figure}

In the top left panel of figure~\ref{fig:PDF_density}, we show $P(\rho)$ at the radius of the solar system, $r = 8$~kpc, comparing the cases of CDM and SIDM Model I.
Although many low-mass subhalos experience core collapse with the SIDM parameters adopted, the probability for Earth to reside in such a collapsed region is so tiny that the effect does not appear as an enhancement in the density distribution $P(\rho)$.
Instead, the suppression of the density profiles for cored higher-mass subhalos at larger radii in the SIDM case, as seen in figure~\ref{fig:density}, yields a more significant effect.
This is because the larger fractional volume associated with these regions is of a more substantial impact.
We also show the SIDM subhalo contributions from different mass ranges as dotted curves.
The smaller subhalos are denser, but their overall contribution to the total density probability is more suppressed.

These findings, however, depend in detail on the SIDM model.
In the top right panel of figure~\ref{fig:PDF_density}, we show $P(\rho)$ for the SIDM models I--V (table~\ref{tab:SIDM models}).
In particular, SIDM Model III enhances the high-density tail of $P(\rho)$ compared to CDM due to the large collapse fraction for massive subhalos (figure~\ref{fig:Bsh}, left panel). This is noteworthy because SIDM Model III can simultaneously explain astrophysical observations on a variety of scales~\cite{Nadler:2023nrd,Zhang:2024ggu,Kong:2024zyw} while affecting $P(\rho)$ in a distinct manner compared to the other SIDM models we study. Thus, there is an interplay between the SIDM cross section at different velocity scales and the local dark matter density, which both direct detection experiments and astrophysical probes of the Milky Way's density profile may be sensitive to.

Finally, the bottom left (bottom right) panel of figure~\ref{fig:PDF_density} compares the probability density function for SIDM Model~I (Model~III) at different radii. In both cases, the mean value of these distributions shifts to lower densities at larger radii, as expected because densities are lower in the outer regions of the host halo. Interestingly, the high-density tails of the distributions become more pronounced at large radii, particularly for Model III.

It will be interesting to refine our model in order to compare these predictions with cosmological zoom-in simulations in future work. For example, our assumption that subhalos' spatial distribution is mass-independent breaks down for subhalos with large infall masses, which sink to the host center more quickly due to dynamical friction \cite{Nagai:2004ac,Nadler:2022dvo}. Incorporating this effect will require explicitly modelling tidal stripping along subhalos' orbits. Furthermore, it will be important to model the effects of central galaxies on SIDM subhalo populations, which can be important for subhalos with small pericenters~\cite{Garrison-Kimmel:2017zes,Nadler:2017dxq}. In SIDM, the disk may further diversify subhalo populations by accelerating core collapse for some systems and disrupting others that have sufficiently large cores. Understanding the extent to which the disruption of cored SIDM subhalos is physical versus numerical will therefore be critical (e.g., see Refs.~\cite{vandenBosch:2016hjf,vandenBosch:2017ynq,vandenBosch:2018tyt,Errani:2022aru}).

\section{Conclusions}
\label{sec:Conclusions}

We have presented a new semi-analytical model, SASHIMI-SIDM, to efficiently predict SIDM subhalo populations down to arbitrarily low masses. 
By combining the structure-formation theory with a parametric model that connects the CDM subhalos to their SIDM counterpart, calibrated using the CDM and SIDM cosmological simulations, SASHIMI-SIDM runs in minutes without being limited by numerical resolution.
Our main findings are as follows.
\begin{itemize}
    \item By taking benchmark SIDM parameters with $\sigma_0 / m_\chi = 147.1$~cm$^{2}$~g$^{-1}$ and $w = 24.33$~km~s$^{-1}$ in eq.~(\ref{eq:dsigmadcostheta}), we compared the distribution of $V_{\rm max}$ (the maximum circular speed) and $r_{\rm max}$ (radius at which $V_{\rm max}$ is reached) with the results of the corresponding cosmological simulations, for the subhalos with masses heavier than $10^8 M_\odot/h$. We found good agreement for both the CDM and SIDM cases (figure~\ref{fig:contour_Vmax_rmax}).
    \item We then included subhalos with masses below the cosmological simulation's resolution limit. For our fiducial SIDM model (Model I), the fraction of core-collapsed subhalos reaches its maximum of $\sim$30\% around subhalo masses of $10^7 M_\odot$, and the core-collapsed fraction decreases towards lower subhalo masses (figure~\ref{fig:SHMF}). 
    \item In the context of SASHIMI-SIDM, this behaviour can be explained by comparing the core-collapse timescale, and its dependence on $V_{\rm max}$ and $r_{\rm max}$ from eq.~(\ref{eq:collapse timescale}), to the age of the subhalo. We also provide an analytic understanding of the core-collapse fraction; see eq.~(\ref{eq:cc_scaling}) and the related discussion. Our predictions explain the behaviour of the core-collapsed fraction for various SIDM models (figure~\ref{fig:Bsh}, left panel).
    \item We studied the implications for dark matter indirect-detection experiments. For our fiducial SIDM model (Model I), the rate of dark matter annihilation is enhanced by a factor of up to $\sim 4$ compared with the host-halo only expectation. This boost is further enhanced for SIDM models that reach larger cross sections at low relative velocities. For example, for Model~II (table~\ref{tab:SIDM models}), the boost factor is as large as a factor of 10 (figure~\ref{fig:Bsh}, right panel).
     \item We discussed the implications for local dark matter densities in the Milky Way. For SIDM Model I, even if a substantial fraction of the subhalos experience core-collapse, the total volume associated with these collapsed subhalos is tiny. Instead, we found the probability of finding a dense subhalo at the location of the solar system was decreased, compared to the CDM case, due to core formation of more massive subhalos (figure~\ref{fig:PDF_density}). This is, however, not the case for other models that predict larger fraction of collapsed subhalos (e.g., Model~III).
\end{itemize}

Our finding that the core-collapse fraction peaks at a particular mass scale set by the underlying SIDM microphysics is particularly striking. For example, in our fiducial SIDM Model~I from refs.~\cite{Turner:2020vlf,Yang:2022mxl}, which is motivated by the diverse rotation curves and inner densities of isolated dwarf galaxies and Milky Way satellites the core-collapse fraction peaks at present-day subhalo masses of $\sim 10^7M_\odot$. It will be interesting to search for these systems in the Milky Way, both as hosts of faint satellite galaxies and through their impact on stellar streams. Meanwhile, SIDM Model III, which is motivated by observations of low-concentration ultra-diffuse galaxies and extremely dense strong gravitational lensing perturbers \cite{Nadler:2023nrd}, the core-collapse fraction peaks at higher subhalo masses of $\sim 10^{9}\text{--}10^{10}M_\odot$. Our analytic calculation of the core-collapsed fraction highlights that this result is sensitive to the assumed mass--concentration relation, which should be studied in SIDM models at very low subhalo masses.

We have demonstrated that SASHIMI-SIDM can rapidly and accurately generate SIDM subhalo populations down to arbitrarily low masses. This will enable more comprehensive and efficient scans of SIDM parameter space, allowing these models to be tested using astronomical data. Relevant observables include dwarf galaxy number counts and density profiles~\cite{Simon:2019ojy,Nadler:2024ims}, strong gravitational lensing flux ratio statistics and gravitational imaging data~\cite{Vegetti:2023mgp}, stellar stream perturbations~\cite{Banik:2018pjp}, and pulsar timing array signals~\cite{Ramani:2020hdo}. These can be probed with current and future facilities such as the James Webb Space Telescope~\cite{Gardner:2006ky}, Euclid~\cite{Euclid:2021icp}, the Vera C.\ Rubin Observatory~\cite{LSST:2008ijt}, and the Nancy Grace Roman Space Telescope~\cite{Spergel:2015sza}.
We plan to pursue these tests of SIDM in future studies.


\acknowledgments

This work was supported by MEXT KAKENHI under grant numbers JP20H05850 (SA) and JP20H05861 (SA and SH), the John Templeton Foundation under grant ID \#61884 and the U.S. Department of Energy under grant number de-sc0008541 (DY and HBY). We thank the Pollica Physics Center, where this research was initiated, for its warm hospitality. The Pollica Physics Center is supported by the Regione Campania, Universit\`a degli Studidi Salerno, Università degli Studi di Napoli ``Federico I,'' the Physics Department ``Ettore Pancini'' and ``E.R.\ Caianiello,'' and the Istituto Nazionale di Fisica Nucleare.


\bibliographystyle{JHEP}
\bibliography{biblio.bib}

\providecommand{\href}[2]{#2}\begingroup\raggedright\begin{thebibliography}{10}

\bibitem{Bullock:2017xww}
J.S.~Bullock and M.~Boylan-Kolchin, \emph{{Small-Scale Challenges to the $\Lambda$CDM Paradigm}}, \href{https://doi.org/10.1146/annurev-astro-091916-055313}{\emph{Ann. Rev. Astron. Astrophys.} {\bfseries 55} (2017) 343} [\href{https://arxiv.org/abs/1707.04256}{{\ttfamily 1707.04256}}].

\bibitem{Nesti:2023tid}
F.~Nesti, P.~Salucci and N.~Turini, \emph{{The Quest for the Nature of the Dark Matter: The Need of a New Paradigm}}, \href{https://doi.org/10.3390/astronomy2020007}{\emph{Astronomy} {\bfseries 2} (2023) 90} [\href{https://arxiv.org/abs/2308.02004}{{\ttfamily 2308.02004}}].

\bibitem{Spergel:1999mh}
D.N.~Spergel and P.J.~Steinhardt, \emph{{Observational evidence for selfinteracting cold dark matter}}, \href{https://doi.org/10.1103/PhysRevLett.84.3760}{\emph{Phys. Rev. Lett.} {\bfseries 84} (2000) 3760} [\href{https://arxiv.org/abs/astro-ph/9909386}{{\ttfamily astro-ph/9909386}}].

\bibitem{Tulin:2017ara}
S.~Tulin and H.-B.~Yu, \emph{{Dark Matter Self-interactions and Small Scale Structure}}, \href{https://doi.org/10.1016/j.physrep.2017.11.004}{\emph{Phys. Rept.} {\bfseries 730} (2018) 1} [\href{https://arxiv.org/abs/1705.02358}{{\ttfamily 1705.02358}}].

\bibitem{Adhikari:2022sbh}
S.~Adhikari et~al., \emph{{Astrophysical Tests of Dark Matter Self-Interactions}},  \href{https://arxiv.org/abs/2207.10638}{{\ttfamily 2207.10638}}.

\bibitem{Navarro:1995iw}
J.F.~Navarro, C.S.~Frenk and S.D.M.~White, \emph{{The Structure of cold dark matter halos}}, \href{https://doi.org/10.1086/177173}{\emph{Astrophys. J.} {\bfseries 462} (1996) 563} [\href{https://arxiv.org/abs/astro-ph/9508025}{{\ttfamily astro-ph/9508025}}].

\bibitem{Balberg:2002ue}
S.~Balberg, S.L.~Shapiro and S.~Inagaki, \emph{{Selfinteracting dark matter halos and the gravothermal catastrophe}}, \href{https://doi.org/10.1086/339038}{\emph{Astrophys. J.} {\bfseries 568} (2002) 475} [\href{https://arxiv.org/abs/astro-ph/0110561}{{\ttfamily astro-ph/0110561}}].

\bibitem{Koda:2011yb}
J.~Koda and P.R.~Shapiro, \emph{{Gravothermal collapse of isolated self-interacting dark matter haloes: N-body simulation versus the fluid model}}, \href{https://doi.org/10.1111/j.1365-2966.2011.18684.x}{\emph{Mon. Not. Roy. Astron. Soc.} {\bfseries 415} (2011) 1125} [\href{https://arxiv.org/abs/1101.3097}{{\ttfamily 1101.3097}}].

\bibitem{Essig:2018pzq}
R.~Essig, S.D.~Mcdermott, H.-B.~Yu and Y.-M.~Zhong, \emph{{Constraining Dissipative Dark Matter Self-Interactions}}, \href{https://doi.org/10.1103/PhysRevLett.123.121102}{\emph{Phys. Rev. Lett.} {\bfseries 123} (2019) 121102} [\href{https://arxiv.org/abs/1809.01144}{{\ttfamily 1809.01144}}].

\bibitem{Kaplinghat:2019svz}
M.~Kaplinghat, M.~Valli and H.-B.~Yu, \emph{{Too Big To Fail in Light of Gaia}}, \href{https://doi.org/10.1093/mnras/stz2511}{\emph{Mon. Not. Roy. Astron. Soc.} {\bfseries 490} (2019) 231} [\href{https://arxiv.org/abs/1904.04939}{{\ttfamily 1904.04939}}].

\bibitem{Nadler:2023nrd}
E.O.~Nadler, D.~Yang and H.-B.~Yu, \emph{{A Self-interacting Dark Matter Solution to the Extreme Diversity of Low-mass Halo Properties}}, \href{https://doi.org/10.3847/2041-8213/ad0e09}{\emph{Astrophys. J. Lett.} {\bfseries 958} (2023) L39} [\href{https://arxiv.org/abs/2306.01830}{{\ttfamily 2306.01830}}].

\bibitem{Kamada:2016euw}
A.~Kamada, M.~Kaplinghat, A.B.~Pace and H.-B.~Yu, \emph{{How the Self-Interacting Dark Matter Model Explains the Diverse Galactic Rotation Curves}}, \href{https://doi.org/10.1103/PhysRevLett.119.111102}{\emph{Phys. Rev. Lett.} {\bfseries 119} (2017) 111102} [\href{https://arxiv.org/abs/1611.02716}{{\ttfamily 1611.02716}}].

\bibitem{Ren:2018jpt}
T.~Ren, A.~Kwa, M.~Kaplinghat and H.-B.~Yu, \emph{{Reconciling the Diversity and Uniformity of Galactic Rotation Curves with Self-Interacting Dark Matter}}, \href{https://doi.org/10.1103/PhysRevX.9.031020}{\emph{Phys. Rev. X} {\bfseries 9} (2019) 031020} [\href{https://arxiv.org/abs/1808.05695}{{\ttfamily 1808.05695}}].

\bibitem{Sameie:2019zfo}
O.~Sameie, H.-B.~Yu, L.V.~Sales, M.~Vogelsberger and J.~Zavala, \emph{{Self-Interacting Dark Matter Subhalos in the Milky Way\textquoteright{}s Tides}}, \href{https://doi.org/10.1103/PhysRevLett.124.141102}{\emph{Phys. Rev. Lett.} {\bfseries 124} (2020) 141102} [\href{https://arxiv.org/abs/1904.07872}{{\ttfamily 1904.07872}}].

\bibitem{Correa:2020qam}
C.A.~Correa, \emph{{Constraining velocity-dependent self-interacting dark matter with the Milky Way\textquoteright{}s dwarf spheroidal galaxies}}, \href{https://doi.org/10.1093/mnras/stab506}{\emph{Mon. Not. Roy. Astron. Soc.} {\bfseries 503} (2021) 920} [\href{https://arxiv.org/abs/2007.02958}{{\ttfamily 2007.02958}}].

\bibitem{Turner:2020vlf}
H.C.~Turner, M.R.~Lovell, J.~Zavala and M.~Vogelsberger, \emph{{The onset of gravothermal core collapse in velocity-dependent self-interacting dark matter subhaloes}}, \href{https://doi.org/10.1093/mnras/stab1725}{\emph{Mon. Not. Roy. Astron. Soc.} {\bfseries 505} (2021) 5327} [\href{https://arxiv.org/abs/2010.02924}{{\ttfamily 2010.02924}}].

\bibitem{Correa:2022dey}
C.A.~Correa, M.~Schaller, S.~Ploeckinger, N.~Anau~Montel, C.~Weniger and S.~Ando, \emph{{TangoSIDM: Tantalizing models of Self-Interacting Dark Matter}}, \href{https://doi.org/10.1093/mnras/stac2830}{\emph{Mon. Not. Roy. Astron. Soc.} {\bfseries 517} (2022) 3045} [\href{https://arxiv.org/abs/2206.11298}{{\ttfamily 2206.11298}}].

\bibitem{Yang:2023jwn}
D.~Yang, E.O.~Nadler, H.-B.~Yu and Y.-M.~Zhong, \emph{{A parametric model for self-interacting dark matter halos}}, \href{https://doi.org/10.1088/1475-7516/2024/02/032}{\emph{JCAP} {\bfseries 02} (2024) 032} [\href{https://arxiv.org/abs/2305.16176}{{\ttfamily 2305.16176}}].

\bibitem{Nishikawa:2019lsc}
H.~Nishikawa, K.K.~Boddy and M.~Kaplinghat, \emph{{Accelerated core collapse in tidally stripped self-interacting dark matter halos}}, \href{https://doi.org/10.1103/PhysRevD.101.063009}{\emph{Phys. Rev. D} {\bfseries 101} (2020) 063009} [\href{https://arxiv.org/abs/1901.00499}{{\ttfamily 1901.00499}}].

\bibitem{Kahlhoefer:2019oyt}
F.~Kahlhoefer, M.~Kaplinghat, T.R.~Slatyer and C.-L.~Wu, \emph{{Diversity in density profiles of self-interacting dark matter satellite halos}}, \href{https://doi.org/10.1088/1475-7516/2019/12/010}{\emph{JCAP} {\bfseries 12} (2019) 010} [\href{https://arxiv.org/abs/1904.10539}{{\ttfamily 1904.10539}}].

\bibitem{Zeng:2021ldo}
Z.C.~Zeng, A.H.G.~Peter, X.~Du, A.~Benson, S.~Kim, F.~Jiang et~al., \emph{{Core-collapse, evaporation, and tidal effects: the life story of a self-interacting dark matter subhalo}}, \href{https://doi.org/10.1093/mnras/stac1094}{\emph{Mon. Not. Roy. Astron. Soc.} {\bfseries 513} (2022) 4845} [\href{https://arxiv.org/abs/2110.00259}{{\ttfamily 2110.00259}}].

\bibitem{Yang:2021kdf}
D.~Yang and H.-B.~Yu, \emph{{Self-interacting dark matter and small-scale gravitational lenses in galaxy clusters}}, \href{https://doi.org/10.1103/PhysRevD.104.103031}{\emph{Phys. Rev. D} {\bfseries 104} (2021) 103031} [\href{https://arxiv.org/abs/2102.02375}{{\ttfamily 2102.02375}}].

\bibitem{Shirasaki:2022ttb}
M.~Shirasaki, T.~Okamoto and S.~Ando, \emph{{Modelling self-interacting dark matter substructures~\textendash{} I.~ Calibration with N-body simulations of a Milky-Way-sized halo and its satellite}}, \href{https://doi.org/10.1093/mnras/stac2539}{\emph{Mon. Not. Roy. Astron. Soc.} {\bfseries 516} (2022) 4594} [\href{https://arxiv.org/abs/2205.09920}{{\ttfamily 2205.09920}}].

\bibitem{Andrade:2023fgr}
K.E.~Andrade, M.~Kaplinghat and M.~Valli, \emph{{Halo densities and pericenter distances of the bright Milky Way satellites as a test of dark matter physics}}, \href{https://doi.org/10.1093/mnras/stae1716}{\emph{Mon. Not. Roy. Astron. Soc.} {\bfseries 532} (2024) 4157} [\href{https://arxiv.org/abs/2311.01528}{{\ttfamily 2311.01528}}].

\bibitem{Meskhidze:2022hwm}
H.~Meskhidze, F.J.~Mercado, O.~Sameie, V.H.~Robles, J.S.~Bullock, M.~Kaplinghat et~al., \emph{{Comparing implementations of self-interacting dark matter in the gizmo and arepo codes}}, \href{https://doi.org/10.1093/mnras/stac1056}{\emph{Mon. Not. Roy. Astron. Soc.} {\bfseries 513} (2022) 2600} [\href{https://arxiv.org/abs/2203.06035}{{\ttfamily 2203.06035}}].

\bibitem{Palubski:2024ibb}
I.~Palubski, O.~Slone, M.~Kaplinghat, M.~Lisanti and F.~Jiang, \emph{{Numerical Challenges in Modeling Gravothermal Collapse in Self-Interacting Dark Matter Halos}},  \href{https://arxiv.org/abs/2402.12452}{{\ttfamily 2402.12452}}.

\bibitem{Fischer:2024eaz}
M.S.~Fischer, K.~Dolag and H.-B.~Yu, \emph{{Numerical challenges for energy conservation in N-body simulations of collapsing self-interacting dark matter haloes}},  \href{https://arxiv.org/abs/2403.00739}{{\ttfamily 2403.00739}}.

\bibitem{Outmezguine:2022bhq}
N.J.~Outmezguine, K.K.~Boddy, S.~Gad-Nasr, M.~Kaplinghat and L.~Sagunski, \emph{{Universal gravothermal evolution of isolated self-interacting dark matter halos for velocity-dependent cross sections}},  \href{https://arxiv.org/abs/2204.06568}{{\ttfamily 2204.06568}}.

\bibitem{Zhong:2023yzk}
Y.-M.~Zhong, D.~Yang and H.-B.~Yu, \emph{{The impact of baryonic potentials on the gravothermal evolution of self-interacting dark matter haloes}},  \href{https://arxiv.org/abs/2306.08028}{{\ttfamily 2306.08028}}.

\bibitem{Yang:2024tba}
D.~Yang, \emph{{Exploring Self-Interacting Dark Matter Halos with Diverse Baryonic Distributions: A Parametric Approach}},  \href{https://arxiv.org/abs/2405.03787}{{\ttfamily 2405.03787}}.

\bibitem{Yang:2022hkm}
D.~Yang and H.-B.~Yu, \emph{{Gravothermal evolution of dark matter halos with differential elastic scattering}}, \href{https://doi.org/10.1088/1475-7516/2022/09/077}{\emph{JCAP} {\bfseries 09} (2022) 077} [\href{https://arxiv.org/abs/2205.03392}{{\ttfamily 2205.03392}}].

\bibitem{Yang:2022zkd}
S.~Yang, X.~Du, Z.C.~Zeng, A.~Benson, F.~Jiang, E.O.~Nadler et~al., \emph{{Gravothermal Solutions of SIDM Halos: Mapping from Constant to Velocity-dependent Cross Section}}, \href{https://doi.org/10.3847/1538-4357/acbd49}{\emph{Astrophys. J.} {\bfseries 946} (2023) 47} [\href{https://arxiv.org/abs/2205.02957}{{\ttfamily 2205.02957}}].

\bibitem{Yang:2024uqb}
D.~Yang, E.O.~Nadler and H.-B.~Yu, \emph{{Testing the parametric model for self-interacting dark matter using matched halos in cosmological simulations}},  \href{https://arxiv.org/abs/2406.10753}{{\ttfamily 2406.10753}}.

\bibitem{Yang:2022mxl}
D.~Yang, E.O.~Nadler and H.-B.~Yu, \emph{{Strong Dark Matter Self-interactions Diversify Halo Populations within and surrounding the Milky Way}}, \href{https://doi.org/10.3847/1538-4357/acc73e}{\emph{Astrophys. J.} {\bfseries 949} (2023) 67} [\href{https://arxiv.org/abs/2211.13768}{{\ttfamily 2211.13768}}].

\bibitem{Hiroshima:2018kfv}
N.~Hiroshima, S.~Ando and T.~Ishiyama, \emph{{Modeling evolution of dark matter substructure and annihilation boost}}, \href{https://doi.org/10.1103/PhysRevD.97.123002}{\emph{Phys. Rev. D} {\bfseries 97} (2018) 123002} [\href{https://arxiv.org/abs/1803.07691}{{\ttfamily 1803.07691}}].

\bibitem{Ando:2020yyk}
S.~Ando, A.~Geringer-Sameth, N.~Hiroshima, S.~Hoof, R.~Trotta and M.G.~Walker, \emph{{Structure formation models weaken limits on WIMP dark matter from dwarf spheroidal galaxies}}, \href{https://doi.org/10.1103/PhysRevD.102.061302}{\emph{Phys. Rev. D} {\bfseries 102} (2020) 061302} [\href{https://arxiv.org/abs/2002.11956}{{\ttfamily 2002.11956}}].

\bibitem{Bond:1990iw}
J.R.~Bond, S.~Cole, G.~Efstathiou and N.~Kaiser, \emph{{Excursion set mass functions for hierarchical Gaussian fluctuations}}, \href{https://doi.org/10.1086/170520}{\emph{Astrophys. J.} {\bfseries 379} (1991) 440}.

\bibitem{Yang:2011rf}
X.~Yang, H.J.~Mo, Y.~Zhang and F.C.v.d.~Bosch, \emph{{An analytical model for the accretion of dark matter subhalos}}, \href{https://doi.org/10.1088/0004-637X/741/1/13}{\emph{Astrophys. J.} {\bfseries 741} (2011) 13} [\href{https://arxiv.org/abs/1104.1757}{{\ttfamily 1104.1757}}].

\bibitem{Hiroshima:2022khy}
N.~Hiroshima, S.~Ando and T.~Ishiyama, \emph{{Semi-analytical frameworks for subhaloes from the smallest to the largest scale}}, \href{https://doi.org/10.1093/mnras/stac2857}{\emph{Mon. Not. Roy. Astron. Soc.} {\bfseries 517} (2022) 2728} [\href{https://arxiv.org/abs/2206.01358}{{\ttfamily 2206.01358}}].

\bibitem{vandenBosch:2004zs}
F.C.~van~den Bosch, G.~Tormen and C.~Giocoli, \emph{{The Mass function and average mass loss rate of dark matter subhaloes}}, \href{https://doi.org/10.1111/j.1365-2966.2005.08964.x}{\emph{Mon. Not. Roy. Astron. Soc.} {\bfseries 359} (2005) 1029} [\href{https://arxiv.org/abs/astro-ph/0409201}{{\ttfamily astro-ph/0409201}}].

\bibitem{Jiang:2014nsa}
F.~Jiang and F.C.~van~den Bosch, \emph{{Statistics of dark matter substructure \textendash{} I. Model and universal fitting functions}}, \href{https://doi.org/10.1093/mnras/stw439}{\emph{Mon. Not. Roy. Astron. Soc.} {\bfseries 458} (2016) 2848} [\href{https://arxiv.org/abs/1403.6827}{{\ttfamily 1403.6827}}].

\bibitem{Correa:2014xma}
C.A.~Correa, J.S.B.~Wyithe, J.~Schaye and A.R.~Duffy, \emph{{The accretion history of dark matter haloes \textendash{} I. The physical origin of the universal function}}, \href{https://doi.org/10.1093/mnras/stv689}{\emph{Mon. Not. Roy. Astron. Soc.} {\bfseries 450} (2015) 1514} [\href{https://arxiv.org/abs/1409.5228}{{\ttfamily 1409.5228}}].

\bibitem{Bryan:1997dn}
G.L.~Bryan and M.L.~Norman, \emph{{Statistical properties of x-ray clusters: Analytic and numerical comparisons}}, \href{https://doi.org/10.1086/305262}{\emph{Astrophys. J.} {\bfseries 495} (1998) 80} [\href{https://arxiv.org/abs/astro-ph/9710107}{{\ttfamily astro-ph/9710107}}].

\bibitem{Correa:2015dva}
C.A.~Correa, J.S.B.~Wyithe, J.~Schaye and A.R.~Duffy, \emph{{The accretion history of dark matter haloes \textendash{} III. A physical model for the concentration\textendash{}mass relation}}, \href{https://doi.org/10.1093/mnras/stv1363}{\emph{Mon. Not. Roy. Astron. Soc.} {\bfseries 452} (2015) 1217} [\href{https://arxiv.org/abs/1502.00391}{{\ttfamily 1502.00391}}].

\bibitem{Ishiyama:2011af}
T.~Ishiyama, J.~Makino, S.~Portegies~Zwart, D.~Groen, K.~Nitadori, S.~Rieder et~al., \emph{{The Cosmogrid Simulation: Statistical Properties of Small Dark Matter Halos}}, \href{https://doi.org/10.1088/0004-637X/767/2/146}{\emph{Astrophys. J.} {\bfseries 767} (2013) 146} [\href{https://arxiv.org/abs/1101.2020}{{\ttfamily 1101.2020}}].

\bibitem{Penarrubia:2010jk}
J.~Penarrubia, A.J.~Benson, M.G.~Walker, G.~Gilmore, A.~McConnachie and L.~Mayer, \emph{{The impact of dark matter cusps and cores on the satellite galaxy population around spiral galaxies}}, \href{https://doi.org/10.1111/j.1365-2966.2010.16762.x}{\emph{Mon. Not. Roy. Astron. Soc.} {\bfseries 406} (2010) 1290} [\href{https://arxiv.org/abs/1002.3376}{{\ttfamily 1002.3376}}].

\bibitem{Nadler:2020ulu}
E.O.~Nadler, A.~Banerjee, S.~Adhikari, Y.-Y.~Mao and R.H.~Wechsler, \emph{{Signatures of Velocity-Dependent Dark Matter Self-Interactions in Milky Way-mass Halos}}, \href{https://doi.org/10.3847/1538-4357/ab94b0}{\emph{Astrophys. J.} {\bfseries 896} (2020) 112} [\href{https://arxiv.org/abs/2001.08754}{{\ttfamily 2001.08754}}].

\bibitem{Pollack:2014rja}
J.~Pollack, D.N.~Spergel and P.J.~Steinhardt, \emph{{Supermassive Black Holes from Ultra-Strongly Self-Interacting Dark Matter}}, \href{https://doi.org/10.1088/0004-637X/804/2/131}{\emph{Astrophys. J.} {\bfseries 804} (2015) 131} [\href{https://arxiv.org/abs/1501.00017}{{\ttfamily 1501.00017}}].

\bibitem{Gilman:2021sdr}
D.~Gilman, J.~Bovy, T.~Treu, A.~Nierenberg, S.~Birrer, A.~Benson et~al., \emph{{Strong lensing signatures of self-interacting dark matter in low-mass haloes}}, \href{https://doi.org/10.1093/mnras/stab2335}{\emph{Mon. Not. Roy. Astron. Soc.} {\bfseries 507} (2021) 2432} [\href{https://arxiv.org/abs/2105.05259}{{\ttfamily 2105.05259}}].

\bibitem{Kaplinghat:2015aga}
M.~Kaplinghat, S.~Tulin and H.-B.~Yu, \emph{{Dark Matter Halos as Particle Colliders: Unified Solution to Small-Scale Structure Puzzles from Dwarfs to Clusters}}, \href{https://doi.org/10.1103/PhysRevLett.116.041302}{\emph{Phys. Rev. Lett.} {\bfseries 116} (2016) 041302} [\href{https://arxiv.org/abs/1508.03339}{{\ttfamily 1508.03339}}].

\bibitem{Andrade:2020lqq}
K.E.~Andrade, J.~Fuson, S.~Gad-Nasr, D.~Kong, Q.~Minor, M.G.~Roberts et~al., \emph{{A stringent upper limit on dark matter self-interaction cross-section from cluster strong lensing}}, \href{https://doi.org/10.1093/mnras/stab3241}{\emph{Mon. Not. Roy. Astron. Soc.} {\bfseries 510} (2021) 54} [\href{https://arxiv.org/abs/2012.06611}{{\ttfamily 2012.06611}}].

\bibitem{Zheng:2023myp}
H.~Zheng, S.~Bose, C.S.~Frenk, L.~Gao, A.~Jenkins, S.~Liao et~al., \emph{{The abundance of dark matter haloes down to Earth mass}},  \href{https://arxiv.org/abs/2310.16093}{{\ttfamily 2310.16093}}.

\bibitem{Shah:2023qcw}
N.~Shah and S.~Adhikari, \emph{{The abundance of core--collapsed subhalos in SIDM: insights from structure formation in $\Lambda$CDM}},  \href{https://arxiv.org/abs/2308.16342}{{\ttfamily 2308.16342}}.

\bibitem{Feng:2020kxv}
W.-X.~Feng, H.-B.~Yu and Y.-M.~Zhong, \emph{{Seeding Supermassive Black Holes with Self-interacting Dark Matter: A Unified Scenario with Baryons}}, \href{https://doi.org/10.3847/2041-8213/ac04b0}{\emph{Astrophys. J. Lett.} {\bfseries 914} (2021) L26} [\href{https://arxiv.org/abs/2010.15132}{{\ttfamily 2010.15132}}].

\bibitem{Tulin:2013teo}
S.~Tulin, H.-B.~Yu and K.M.~Zurek, \emph{{Beyond Collisionless Dark Matter: Particle Physics Dynamics for Dark Matter Halo Structure}}, \href{https://doi.org/10.1103/PhysRevD.87.115007}{\emph{Phys. Rev. D} {\bfseries 87} (2013) 115007} [\href{https://arxiv.org/abs/1302.3898}{{\ttfamily 1302.3898}}].

\bibitem{Wu:2022wzw}
Y.~Wu, S.~Baum, K.~Freese, L.~Visinelli and H.-B.~Yu, \emph{{Dark stars powered by self-interacting dark matter}}, \href{https://doi.org/10.1103/PhysRevD.106.043028}{\emph{Phys. Rev. D} {\bfseries 106} (2022) 043028} [\href{https://arxiv.org/abs/2205.10904}{{\ttfamily 2205.10904}}].

\bibitem{Kaplinghat:2013xca}
M.~Kaplinghat, R.E.~Keeley, T.~Linden and H.-B.~Yu, \emph{{Tying Dark Matter to Baryons with Self-interactions}}, \href{https://doi.org/10.1103/PhysRevLett.113.021302}{\emph{Phys. Rev. Lett.} {\bfseries 113} (2014) 021302} [\href{https://arxiv.org/abs/1311.6524}{{\ttfamily 1311.6524}}].

\bibitem{Sameie:2018chj}
O.~Sameie, P.~Creasey, H.-B.~Yu, L.V.~Sales, M.~Vogelsberger and J.~Zavala, \emph{{The impact of baryonic discs on the shapes and profiles of self-interacting dark matter haloes}}, \href{https://doi.org/10.1093/mnras/sty1516}{\emph{Mon. Not. Roy. Astron. Soc.} {\bfseries 479} (2018) 359} [\href{https://arxiv.org/abs/1801.09682}{{\ttfamily 1801.09682}}].

\bibitem{Robles:2019mfq}
V.H.~Robles, T.~Kelley, J.S.~Bullock and M.~Kaplinghat, \emph{{The Milky Way\textquoteright{}s halo and subhaloes in self-interacting dark matter}}, \href{https://doi.org/10.1093/mnras/stz2345}{\emph{Mon. Not. Roy. Astron. Soc.} {\bfseries 490} (2019) 2117} [\href{https://arxiv.org/abs/1903.01469}{{\ttfamily 1903.01469}}].

\bibitem{Rose:2022mqj}
J.C.~Rose, P.~Torrey, M.~Vogelsberger and S.~O'Neil, \emph{{Unravelling the interplay between SIDM and baryons in MW haloes: defining where baryons dictate heat transfer}}, \href{https://doi.org/10.1093/mnras/stac3634}{\emph{Mon. Not. Roy. Astron. Soc.} {\bfseries 519} (2023) 5623} [\href{https://arxiv.org/abs/2206.14830}{{\ttfamily 2206.14830}}].

\bibitem{Ando:2019xlm}
S.~Ando, T.~Ishiyama and N.~Hiroshima, \emph{{Halo Substructure Boosts to the Signatures of Dark Matter Annihilation}}, \href{https://doi.org/10.3390/galaxies7030068}{\emph{Galaxies} {\bfseries 7} (2019) 68} [\href{https://arxiv.org/abs/1903.11427}{{\ttfamily 1903.11427}}].

\bibitem{Horigome:2022gge}
S.~Horigome, K.~Hayashi and S.~Ando, \emph{{Cosmological prior for the J-factor estimation of dwarf spheroidal galaxies}}, \href{https://doi.org/10.1103/PhysRevD.108.083530}{\emph{Phys. Rev. D} {\bfseries 108} (2023) 083530} [\href{https://arxiv.org/abs/2207.10378}{{\ttfamily 2207.10378}}].

\bibitem{Kamionkowski:2008vw}
M.~Kamionkowski and S.M.~Koushiappas, \emph{{Galactic substructure and direct detection of dark matter}}, \href{https://doi.org/10.1103/PhysRevD.77.103509}{\emph{Phys. Rev. D} {\bfseries 77} (2008) 103509} [\href{https://arxiv.org/abs/0801.3269}{{\ttfamily 0801.3269}}].

\bibitem{Kamionkowski:2010mi}
M.~Kamionkowski, S.M.~Koushiappas and M.~Kuhlen, \emph{{Galactic Substructure and Dark Matter Annihilation in the Milky Way Halo}}, \href{https://doi.org/10.1103/PhysRevD.81.043532}{\emph{Phys. Rev. D} {\bfseries 81} (2010) 043532} [\href{https://arxiv.org/abs/1001.3144}{{\ttfamily 1001.3144}}].

\bibitem{Ibarra:2019jac}
A.~Ibarra, B.J.~Kavanagh and A.~Rappelt, \emph{{Impact of substructure on local dark matter searches}}, \href{https://doi.org/10.1088/1475-7516/2019/12/013}{\emph{JCAP} {\bfseries 12} (2019) 013} [\href{https://arxiv.org/abs/1908.00747}{{\ttfamily 1908.00747}}].

\bibitem{Springel:2008cc}
V.~Springel, J.~Wang, M.~Vogelsberger, A.~Ludlow, A.~Jenkins, A.~Helmi et~al., \emph{{The Aquarius Project: the subhalos of galactic halos}}, \href{https://doi.org/10.1111/j.1365-2966.2008.14066.x}{\emph{Mon. Not. Roy. Astron. Soc.} {\bfseries 391} (2008) 1685} [\href{https://arxiv.org/abs/0809.0898}{{\ttfamily 0809.0898}}].

\bibitem{Zhang:2024ggu}
X.~Zhang, H.-B.~Yu, D.~Yang and H.~An, \emph{{Self-interacting dark matter interpretation of Crater II}},  \href{https://arxiv.org/abs/2401.04985}{{\ttfamily 2401.04985}}.

\bibitem{Kong:2024zyw}
D.~Kong, D.~Yang and H.-B.~Yu, \emph{{CDM and SIDM Interpretations of the Strong Gravitational Lensing Object JWST-ER1}},  \href{https://arxiv.org/abs/2402.15840}{{\ttfamily 2402.15840}}.

\bibitem{Nagai:2004ac}
D.~Nagai and A.V.~Kravtsov, \emph{{The radial distribution of galaxies in lcdm clusters}}, \href{https://doi.org/10.1086/426016}{\emph{Astrophys. J.} {\bfseries 618} (2005) 557} [\href{https://arxiv.org/abs/astro-ph/0408273}{{\ttfamily astro-ph/0408273}}].

\bibitem{Nadler:2022dvo}
E.O.~Nadler et~al., \emph{{Symphony: Cosmological Zoom-in Simulation Suites over Four Decades of Host Halo Mass}}, \href{https://doi.org/10.3847/1538-4357/acb68c}{\emph{Astrophys. J.} {\bfseries 945} (2023) 159} [\href{https://arxiv.org/abs/2209.02675}{{\ttfamily 2209.02675}}].

\bibitem{Garrison-Kimmel:2017zes}
S.~Garrison-Kimmel et~al., \emph{{Not so lumpy after all: modelling the depletion of dark matter subhaloes by Milky Way-like galaxies}}, \href{https://doi.org/10.1093/mnras/stx1710}{\emph{Mon. Not. Roy. Astron. Soc.} {\bfseries 471} (2017) 1709} [\href{https://arxiv.org/abs/1701.03792}{{\ttfamily 1701.03792}}].

\bibitem{Nadler:2017dxq}
E.O.~Nadler, Y.-Y.~Mao, R.H.~Wechsler, S.~Garrison-Kimmel and A.~Wetzel, \emph{{Modeling the Impact of Baryons on Subhalo Populations with Machine Learning}}, \href{https://doi.org/10.3847/1538-4357/aac266}{\emph{Astrophys. J.} {\bfseries 859} (2018) 129} [\href{https://arxiv.org/abs/1712.04467}{{\ttfamily 1712.04467}}].

\bibitem{vandenBosch:2016hjf}
F.C.~van~den Bosch, \emph{{Dissecting the Evolution of Dark Matter Subhaloes in the Bolshoi Simulation}}, \href{https://doi.org/10.1093/mnras/stx520}{\emph{Mon. Not. Roy. Astron. Soc.} {\bfseries 468} (2017) 885} [\href{https://arxiv.org/abs/1611.02657}{{\ttfamily 1611.02657}}].

\bibitem{vandenBosch:2017ynq}
F.C.~van~den Bosch, G.~Ogiya, O.~Hahn and A.~Burkert, \emph{{Disruption of Dark Matter Substructure: Fact or Fiction?}}, \href{https://doi.org/10.1093/mnras/stx2956}{\emph{Mon. Not. Roy. Astron. Soc.} {\bfseries 474} (2018) 3043} [\href{https://arxiv.org/abs/1711.05276}{{\ttfamily 1711.05276}}].

\bibitem{vandenBosch:2018tyt}
F.C.~van~den Bosch and G.~Ogiya, \emph{{Dark Matter Substructure in Numerical Simulations: A Tale of Discreteness Noise, Runaway Instabilities, and Artificial Disruption}}, \href{https://doi.org/10.1093/mnras/sty084}{\emph{Mon. Not. Roy. Astron. Soc.} {\bfseries 475} (2018) 4066} [\href{https://arxiv.org/abs/1801.05427}{{\ttfamily 1801.05427}}].

\bibitem{Errani:2022aru}
R.~Errani, J.F.~Navarro, J.~Pe\~narrubia, B.~Famaey and R.~Ibata, \emph{{Dark matter halo cores and the tidal survival of Milky Way satellites}}, \href{https://doi.org/10.1093/mnras/stac3499}{\emph{Mon. Not. Roy. Astron. Soc.} {\bfseries 519} (2022) 384} [\href{https://arxiv.org/abs/2210.01131}{{\ttfamily 2210.01131}}].

\bibitem{Simon:2019ojy}
J.D.~Simon et~al., \emph{{Dynamical Masses for a Complete Census of Local Dwarf Galaxies}},  \href{https://arxiv.org/abs/1903.04743}{{\ttfamily 1903.04743}}.

\bibitem{Nadler:2024ims}
E.O.~Nadler, V.~Gluscevic, T.~Driskell, R.H.~Wechsler, L.A.~Moustakas, A.~Benson et~al., \emph{{Forecasts for Galaxy Formation and Dark Matter Constraints from Dwarf Galaxy Surveys}},  \href{https://arxiv.org/abs/2401.10318}{{\ttfamily 2401.10318}}.

\bibitem{Vegetti:2023mgp}
S.~Vegetti et~al., \emph{{Strong gravitational lensing as a probe of dark matter}},  \href{https://arxiv.org/abs/2306.11781}{{\ttfamily 2306.11781}}.

\bibitem{Banik:2018pjp}
N.~Banik, G.~Bertone, J.~Bovy and N.~Bozorgnia, \emph{{Probing the nature of dark matter particles with stellar streams}}, \href{https://doi.org/10.1088/1475-7516/2018/07/061}{\emph{JCAP} {\bfseries 07} (2018) 061} [\href{https://arxiv.org/abs/1804.04384}{{\ttfamily 1804.04384}}].

\bibitem{Ramani:2020hdo}
H.~Ramani, T.~Trickle and K.M.~Zurek, \emph{{Observability of Dark Matter Substructure with Pulsar Timing Correlations}}, \href{https://doi.org/10.1088/1475-7516/2020/12/033}{\emph{JCAP} {\bfseries 12} (2020) 033} [\href{https://arxiv.org/abs/2005.03030}{{\ttfamily 2005.03030}}].

\bibitem{Gardner:2006ky}
J.P.~Gardner et~al., \emph{{The James Webb Space Telescope}}, \href{https://doi.org/10.1007/s11214-006-8315-7}{\emph{Space Sci. Rev.} {\bfseries 123} (2006) 485} [\href{https://arxiv.org/abs/astro-ph/0606175}{{\ttfamily astro-ph/0606175}}].

\bibitem{Euclid:2021icp}
{\scshape Euclid} collaboration, \emph{{Euclid preparation. I. The Euclid Wide Survey}}, \href{https://doi.org/10.1051/0004-6361/202141938}{\emph{Astron. Astrophys.} {\bfseries 662} (2022) A112} [\href{https://arxiv.org/abs/2108.01201}{{\ttfamily 2108.01201}}].

\bibitem{LSST:2008ijt}
{\scshape LSST} collaboration, \emph{{LSST: from Science Drivers to Reference Design and Anticipated Data Products}}, \href{https://doi.org/10.3847/1538-4357/ab042c}{\emph{Astrophys. J.} {\bfseries 873} (2019) 111} [\href{https://arxiv.org/abs/0805.2366}{{\ttfamily 0805.2366}}].

\bibitem{Spergel:2015sza}
D.~Spergel et~al., \emph{{Wide-Field InfrarRed Survey Telescope-Astrophysics Focused Telescope Assets WFIRST-AFTA 2015 Report}},  \href{https://arxiv.org/abs/1503.03757}{{\ttfamily 1503.03757}}.

\end{thebibliography}\endgroup

\end{document}